\documentclass[%
 reprint,
superscriptaddress,
reprint,
 amsmath,amssymb,
 aps,
prb,
floatfix,
dvipsnames, 9pt
]{revtex4-2}
\usepackage{array}[=2016-10-06]
\usepackage{graphicx}
\usepackage{dcolumn}
\usepackage{bm}
 \usepackage{hyperref}
\usepackage{tikz}
\usetikzlibrary{arrows}
\usetikzlibrary{calc}
\usepackage{enumerate}
\usepackage{caption}
\usepackage{ragged2e}
\usetikzlibrary{3d, decorations.pathreplacing}
\usepackage{float}
\usepackage{subcaption}
\DeclareMathAlphabet{\mathpzc}{OT1}{pzc}{m}{it}
\usepackage{pgfplots}
\usepackage[export]{adjustbox}
\usepackage{float}
\usepackage{xcolor}
\usepackage{soul,ulem}

\usepackage{upgreek}

\usetikzlibrary{positioning}
\definecolor{maroon}{cmyk}{0,0.87,0.48,0.32}
\definecolor{Gray}{gray}{0.85}
\definecolor{LightCyan}{rgb}{0.88,1,1}
\definecolor{babyblueeyes}{rgb}{0.7, 0.84, 0.96} 
\definecolor{bluegray}{rgb}{0.4, 0.6, 0.9}
\definecolor{bluegraym}{rgb}{0.4, 0.6, 0.99}
\definecolor{airforceblue}{rgb}{0.0, 0.53, 0.74}

\newcommand \figref{{\bf \color{blue}Fig.}~\ref}
\newcommand \tabref{{\bf \color{blue}Table.}~\ref}
\makeatletter
\let\@msm@th@eqref\eqref
\hypersetup{
    colorlinks=true, 
    linkcolor=blue,  
    citecolor=red,   
    urlcolor=cyan   
}
\renewcommand{\eqref}[1]{%
    \begingroup
    \leavevmode
    \color{blue}%
    \hypersetup{linkbordercolor=[named]{blue}, linkcolor=[named]{blue}}%
    \@msm@th@eqref{#1}%
    \endgroup
}

\begin{document}

\title{{Distinct Spatiotemporal Dynamics of Thermoelectric Transport Across Superconducting Transition}}


\author{Rajae Malek}
    \email[Correspondence email address: ]{malekrajae@gmail.com}
\affiliation{Shanghai Research Center for Quantum Sciences, Shanghai 201315, China}
\affiliation{Tsung-Dao Lee Institute \& School of Physics and Astronomy,
Shanghai Jiao Tong University, Shanghai, 212, China}

\author{Qing-Dong Jiang}
\email[Correspondence email address: ]{qingdong.jiang@sjtu.edu.cn}
\affiliation{Shanghai Research Center for Quantum Sciences, Shanghai 201315, China}
\affiliation{Tsung-Dao Lee Institute \& School of Physics and Astronomy,
Shanghai Jiao Tong University, Shanghai, 212, China}

\author{Haiwen Liu}
\email[Correspondence email address: ]{haiwen.liu@bnu.edu.cn}
\affiliation{Center for Advanced Quantum Studies, School of Physics and Astronomy, Beijing Normal University, Beijing 100875, China}
\affiliation{Key Laboratory of Multiscale Spin Physics (Ministry of Education), Beijing Normal University, Beijing 100875, China}

\date{\today}

\begin{abstract}

We investigate the relaxation dynamics of heat transport in superconductors, shaped by the interplay of diffusion, nonlinearity, and magnetic fields. Focusing on regimes near the critical temperature $T_c$, we analyze two classes of relaxation–diffusion equations that give rise to qualitatively distinct dynamics, which we denote as Type-I (linear) and Type-II (nonlinear). Type-I relaxation, characteristic of the normal state above $T_c$, results in a steady and spatially uniform heat current governed by linear diffusion. By contrast, Type-II relaxation, relevant below $T_c$, exhibits non-steady dynamics marked by pronounced spatial inhomogeneities and an evolving pattern, in which an initially localized hot spot propagates transiently through the system. The striking distinction between these regimes underscores a fundamental shift in transport mechanisms across the superconducting phase transition and provides experimentally relevant predictions in light of emerging techniques for probing local dissipation.

\end{abstract}

 \maketitle
  \onecolumngrid
\section{Introduction}
To understand the nature of superconductivity, researchers have extensively investigated various related phenomena, with superconducting fluctuations being of particular importance and providing crucial insights into the behavior of superconducting systems through decades of study \cite{Var}. These fluctuations, especially near the superconducting transition temperature $T_c$, can significantly influence the properties of materials, particularly in how they respond to external fields and thermal perturbations \cite{Sco}. Landmark contributions in this regard were provided by Aslamazov and Larkin \cite{Asla} and Maki  \cite{maki1968critical}, who demonstrated that fluctuations in the normal state of superconductors lead to a dramatic divergence in electrical conductivity as the system approaches $T_c$. This observation paved the way for further exploration of how fluctuations at the critical temperature can impact transport properties like electric and heat conductivity, providing new perspectives on the behavior of superconducting systems near phase transitions \cite{huebener1995superconductors}. Building on these early insights, the study of superconducting fluctuations has evolved to incorporate advanced methods from both the phenomenological Ginzburg-Landau (GL) theory and microscopic approaches \cite{Lar, Var}. These fluctuations are key factors in determining experimentally observable quantities such as electrical conductivity, heat current, and the Nernst coefficient \cite{behnia2015fundamentals}. The Nernst effect has been identified as an excellent experimental probe for studying the response of superconducting fluctuations near $T_c$ under thermal gradients and magnetic fields \cite{pourret2009nernst, behnia2016nernst, ullah1990critical, Ussishkin2002PRL, vishwanath2008nernst, wang2006nernst, cyr2018pseudogap, hu2024vortex}.{\color{black}  While steady-state transport properties near $T_c$ have been extensively studied, the underlying spatiotemporal dynamics governing how heat propagates and energy dissipates across the transition remain less understood. Previous studies have focused primarily on spatially averaged transport coefficients, which do not provide direct insight into the real-time evolution and local distribution of heat flow. Yet, a theoretical framework that directly contrasts the spatiotemporal signatures of linear (above $T_c$) and nonlinear (below $T_c$) relaxation has been lacking.} \vskip 2pt

Our work leverages the time-dependent Ginzburg-Landau (TDGL) framework to delve deeper into the dynamics of superconducting fluctuations, particularly the relaxation processes that govern the evolution of the superconducting order parameter. The TDGL framework allows us to explore how the superconducting state responds to thermal fluctuations and external perturbations, providing a versatile tool for understanding both linear and nonlinear effects in these systems \cite{Sco,schmid1966time, binder1973time}. A central aspect of this work is the classification of relaxation processes near $T_c$ into two distinct regimes: Type-I and Type-II relaxation.
{\color{black}Type-I relaxation, typically occurring above $T_c$ \cite{schmid1966time}, describes dynamics governed by a linear relaxation-diffusion equation, characterized by the slow relaxation of superconducting fluctuations within the normal state. In contrast, Type-II relaxation, relevant below $T_c$ \cite{tang1995time}, involves inherently nonlinear dynamics (often referred to as ``quenched" dynamics). In this regime, the interplay between thermal fluctuations, magnetic fields, and, crucially, the nonlinear self-interaction of the order parameter dictates the system's evolution.} Thus, the relaxation dynamics of superconducting systems are inherently complex, involving the interplay of dissipation, coherence, and nonlinear effects.\cite{schmidt1968onset, tang1995time}. The TDGL framework provides a powerful and flexible approach to modeling the evolution of propagators, correlations, and physical quantities such as heat current for both Type-I and Type-II relaxations, capturing the interplay between dissipation and coherence, and handling nonequilibrium dynamics under magnetic field\cite{rosenstein2010ginzburg,ullah1990critical}. By solving these equations, either numerically or analytically, we can gain valuable insights into the dynamics of superconducting systems and their response to external perturbations. The thermoelectric Nernst coefficient plays a critical role in distinguishing between these two relaxation processes \cite{behnia2016nernst}. The sharp difference of heat propagation between Type-I and Type-II relaxations provides experimentalists with a measurable quantity that exhibits distinctive signatures, offering a direct connection between the theoretical understanding of these processes and experimental observations. \vskip 2pt 

{\color{black} Recent experimental and theoretical advances have highlighted the complex interplay between superconducting fluctuations, vortex dynamics, and thermoelectric transport in low-dimensional systems. In particular, studies of fluctuation-driven Nernst response and vortex entropy have provided direct insight into dissipative mechanisms near the superconducting transition \cite{Hu2024}. At the same time, nanoscale and mesoscopic systems have revealed how confinement and geometry can significantly modify thermoelectric transport and vortex behavior \cite{Bouteiller2025,Shokri2025}. More broadly, recent works have explored unconventional transport regimes and nonequilibrium effects beyond the standard Gaussian fluctuation framework \cite{Wu2024}. These developments underline the importance of approaches that go beyond steady-state descriptions and capture both spatial and temporal evolution of transport phenomena. In this context, our work complements these studies by focusing on the spatiotemporal dynamics of heat transport and by explicitly contrasting linear and nonlinear relaxation regimes, thereby providing new insight into transient and spatially resolved thermoelectric behavior.}

In this work, we examine thermoelectric transport across the superconducting transition.  The focus of this study is the spatiotemporal dynamics of superconducting relaxation across the transition. We analyze both Type-I (linear) and Type-II (nonlinear) regimes by examining the time evolution, spatial distribution,  and resulting heat current of the order-parameter propagator. {\color{black}Our results reveal a pronounced contrast between the two dynamical regimes. While the Type-I regime ($T>T_c$) eventually reaches a steady, spatially uniform heat transport consistent with known results, our analysis provides novel insights into its transient diffusive dynamics. Crucially, the Type-II regime ($T<T_c$) exhibits fundamentally different behavior: non-steady dynamics marked by transient amplification, spatially inhomogeneous patterns, and evolving hot spots. This qualitative distinction in the spatiotemporal response constitutes the central prediction of this work, offering clear measurable signatures across the transition. }
Using the $\delta_{-}$Ziti method \cite{rajeq,Aa,Ks}, we numerically solve the linear propagator equation Eq.~\eqref{rdmf} and the reduced nonlinear response equation Eq.~\eqref{type2}, from which the corresponding physical observables, including the heat current, are obtained. This framework reveals the distinct transport characteristics of the two regimes, capturing both linear and nonlinear dynamics in superconducting systems. \vskip 2pt

The outline of this paper is as follows:  In Section \ref{thapp}, we discuss our theoretical results, focusing on the analytical solution of the Type-I relaxation equation, along with the related correlations and heat current in the presence of an electromagnetic field. In section \ref{linsec}, we present our numerical results for solving the linear Type-I equation, while section \ref{nonlinsec} focuses on our numerical contribution to solving the nonlinear Type-II relaxation equation. We conclude with {brief} remarks and perspectives in section \ref{conc}. Appendix \ref{numapp} outlines the key steps in constructing the numerical method $\delta_{-}$Ziti, which is designed to deal with both Type-I and Type-II relaxation equations, which can successfully tackle the analytically intractable nonlinear dynamics inherent in the Type-II regime.

\section{Theoretical results of Type-I relaxation equation}\label{thapp}

We consider two distinct regimes: the Type-I relaxation equation, relevant for temperatures above $T_c$, and the Type-II quenched dynamical relaxation, characterizing the regime below $T_c$. In this section, we present the analytical treatment of the Type-I case. The corresponding numerical analyses are reported in Sections \ref{linsec} and \ref{nonlinsec}:
for Type-I, the numerical simulations are benchmarked against the analytical solutions derived here, while for Type-II—where no closed-form analytical solution exists—numerical simulations provide essential support for our physical interpretations.

\subsection{The Type-I relaxation equation with  magnetic field.}
We begin by considering the Type-I relaxation–diffusion equation, which governs the dynamical evolution of a conventional superconducting system subjected to an external magnetic field in the regime just above the superconducting transition temperature $T_c$. The corresponding propagator, from which the system’s physical properties can be derived, is defined by the following equation:
\cite{Sco,rosenstein2010ginzburg}:
\begin{align}
\begin{split}
   	&\left[ \frac{\partial}{\partial t} - D_{\varphi}\left(\hbar \nabla -2ie \vec{A}\right)^2  + \frac{1}{\tau_s}\right] R_0(\vec{r},t;\vec{r_0},t_0)  =\delta(\vec{r}-\vec{r_0}) \delta(t-t_0).	 
\end{split}
\label{rdmf}
\end{align}
Here, $R_0(\vec{r},t;\vec{r}_0,t_0)$ denotes the propagator of the order parameter, and $D_\varphi$ is the  coefficient describing  the spatial diffusion of superconducting fluctuations. The term $\frac{1}{\tau_s}$ represents the Ginzburg-Landau relaxation rate that controls the temporal decay of the order parameter, so that $\tau_s$ defines the intrinsic relaxation time of the system. The differential operator $(\hbar\nabla - 2ie\vec{A})^2$ describes the kinetic energy of superconducting fluctuations and their coupling to the magnetic vector potential $\vec{A}(\vec{r})$. Thus, Eq. \eqref{rdmf} represents the linearized  time-dependent Ginzburg-Landau equation governing diffusive relaxation above $T_c$ in the presence of electromagnetic fields.\vskip 2pt 
Let $R_0(\vec{r},\vec{r_0})$ be the Fourier transform of the real time propagator $R_0(\vec{r},t;\vec{r_0},t_0)$:
\begin{equation*}
	R_0(\vec{r},t;\vec{r_0},t_0)\equiv \frac{1}{2\pi}\int_{-\infty}^{\infty} e^{i \omega (t-t_0)} R_0(\vec{r},\vec{r_0})\  d\omega,
	\label{fourierdef}
\end{equation*}
where the propagator in frequency space satisfies, 
\begin{equation*}
	\left[  i \omega - D_{\varphi}\left(\hbar \nabla -2ie \vec{A}\right)^2  + \frac{1}{\tau_s}\right] R_0(\vec{r},\vec{r_0})=\delta(\vec{r}-\vec{r_0}).
	\label{fourier}
\end{equation*}
We aim to solve Eq.~\eqref{rdmf} by Hermite polynomials by expanding $R_0(\vec{r},\vec{r_0})=\underset{n, p_x}{\sum} C(n, p_x) e^{i \frac{p_x.x}{\hbar}} \chi_n(y)$. 
{\color{black}
For the sake of simplicity, we adopt the Landau gauge for a uniform magnetic field perpendicular to the $(x,y)$ plane,
\begin{equation*}
\vec{A}=(-By,0,0),
\qquad
\vec{B}=\nabla\times\vec{A}=B\hat{z}.
\end{equation*}
The magnetic field is therefore perpendicular to the film. Under the gauge transformation
\begin{equation*}
\vec{A}\rightarrow \vec{A}+\nabla\chi,
\qquad
\Phi\rightarrow \Phi-\frac{1}{c}\partial_t\chi,
\end{equation*}
the two-point response acquires the standard phase factor, while the covariant derivatives entering the microscopic heat-current operator cancel this phase at coincident points. The local heat-current density constructed in Sec.~II is therefore gauge invariant. In Sec.~IV, by contrast, we use a reduced phenomenological nonlinear model in a fixed Landau gauge for the same perpendicular-field geometry; it is not presented as a full gauge-complete replacement of the complex TDGL theory.} {\color{black}
For notational consistency, throughout the TDGL derivations we use Gaussian units and set $c=1$ in the covariant derivatives unless $c$ is written explicitly to facilitate comparison with standard transport formulas. With this convention, the Cooper-pair magnetic length is $l_B=\sqrt{\hbar/(2eB)}$, and the corresponding dimensionless magnetic term takes the form $(\partial_{\tilde x}+i\tilde y)^2$.} In this case, the solution of Eq.~\eqref{rdmf} is similar to the Landau Level solution of 2D electron gas, with  $\omega_B=  4 e B D_{\varphi}$, $a_B = \sqrt{\frac{\hbar}{2e B}}$ and $y_1=  \frac{p_x}{2 e B}$. Recall that the Hermite functions $\chi_n(y)$ are given by
\begin{equation*}
\chi_n(y):=\frac{1}{\pi^{1/4} a_B^{1/2}\sqrt{2^n n!}}
\exp\left[-\frac{(y-y_1)^2}{2a_B^2}\right]
H_n\left(\frac{y-y_1}{a_B}\right).
\end{equation*} It remains to identify the coefficients $C(p_x,n)$ based on the Hermite polynomial properties. Indeed:
{\small \begin{align*}
	&\left[  i \omega - D_{\varphi}\left(\hbar \nabla -2ie \vec{A}\right)^2  + \frac{1}{\tau_s}\right] R_0(\vec{r},\vec{r_0})\\
	&	= \sum_{n, p_x} C(n, p_x) e^{i \frac{p_x.x}{\hbar}} \left[  i \omega + \frac{1}{\tau_s}+(n+\frac{1}{2}) \hbar \omega_B\right]\chi_n(y)\\
	&= \delta(x-x_0) \delta(y-y_0).
\end{align*}}
Thus, we have the expression of the coefficient $C(n, p_x)$:
{\small \begin{equation*}
		C(n, p_x)= \frac{1}{\sqrt{2^n. n!}}e^{-i p_x \frac{x_0}{\hbar}} . \frac{H_n(\frac{y_0-y_1}{a_B}). \exp(-\frac{(y_0-y_1)^2}{2a_B^2})}{  i \omega + \frac{1}{\tau_s}+(n+\frac{1}{2}) \hbar \omega_B}
	\end{equation*}}
and then the Fourier transform of the propagator $R_0(\vec{r},\vec{r_0})= \underset{n, p_x}{\sum} C(n, p_x) e^{i \frac{p_x.x}{\hbar}} \chi_n(y)$ reads as follows:

  \begin{align}
	&R_0(\vec{r},t;\vec{r_0},t_0) =\frac{1}{2\pi}\int_{-\infty}^{\infty}d\omega e^{i \omega (t-t_0)}. \sum_{ p_x} \sqrt{\frac{\omega_B}{2\pi \hbar D_\phi}}e^{i p_x \frac{x-x_0}{\hbar}}\times\exp\left(-\frac{(y-y_1)^2+(y_0-y_1)^2}{2a_B^2}\right) \nonumber	\\ 
   &\hspace{3.2cm}\times \sum_{n}\frac{1}{2^n. n!} \frac{H_n(\frac{y-y_1}{a_B}). H_n(\frac{y_0-y_1}{a_B})}{  i \omega + \frac{1}{\tau_s}+(n+\frac{1}{2}) \hbar \omega_B}.
	\label{R0}
\end{align}

\subsection{The relaxation and correlation with thermal fluctuations}

In the following, we consider the relaxation–diffusion equation in the presence of thermal noise, denoted by $\xi(\vec{r},t)$
\begin{equation}
	 \frac{\partial}{\partial t} \varphi(\vec{r},t)- D_{\varphi}\left(\hbar \nabla -2ie \vec{A}\right)^2\varphi(\vec{r},t)  + \frac{1}{\tau_s} \varphi(\vec{r},t) =\xi(\vec{r},t)
	\label{neweq}
\end{equation}
where the noise term $\xi(\vec{r},t)$ satisfies local correlation:
\begin{equation*}
\langle \xi^*(\vec{r},t)\, \xi(\vec{r}',t') \rangle
= k_B T\, \delta(\vec{r}-\vec{r}')\,\delta(t-t').
\end{equation*}
When an electric field $\vec{E}$ is applied, it can be described by a scalar potential $ \Phi(\vec{r}, t)= - \vec{E} \cdot \vec{r} e^{-i \omega t} $. Before proceeding, it is important to clarify the notion of {\it{steady state}} used in the sequel of this work. In the Type-I regime ($T>T_c$), it refers to the stationary solution of the linearized time-dependent Ginzburg-Landau (TDGL) equation under a weak applied electric field. The introduced Eq.~\eqref{neweq} describes a driven-dissipative system where the temporal derivative of the order parameter vanishes after the transition dynamics have relaxed, i.e $\partial_t \varphi=0$ in the long time limit. The perturbative treatment of the scalar potential $\Phi(\vec{r},t) = -\vec{E}\ \cdot\!\vec{r}\, e^{-i\omega t}$ and the limit $\omega \to 0$ correspond to the linear-response steady state, in which the energy injected by the external field is balanced by dissipation. 
Thus,  Eq.~\eqref{neweq}  reads:
\begin{equation*}
{\color{black}\left(\frac{\partial}{\partial t} + i \frac{2e}{\hbar c} \Phi(\vec{r}, t)\right)} \varphi(\vec{r},t)- D_{\varphi}\left(\hbar \nabla -2ie \vec{A}\right)^2\varphi(\vec{r},t)  + \frac{1}{\tau_s} \varphi(\vec{r},t) =\xi(\vec{r},t)
\end{equation*}
{The solution of the driven equation can be written as the convolution between the propagator and the noise function:}
\begin{equation*}
	\varphi(\vec{r}, t) = \int R(\vec{r},t;\vec{r'},t') \xi(\vec{r'},t') d\vec{r'} dt'
\end{equation*}
and thus the correlation function can be expressed in terms of propagators:
\begin{align}
	C(\vec{r},t;\vec{r}',t') &=\langle\varphi(\vec{r},t) \varphi^{*}(\vec{r}',t')\rangle \nonumber\\
	&\hspace{0.2cm}= k_B T \int R(\vec{r},t;\vec{r_0},t_0) R^{*}(\vec{r'},t';\vec{r_0},t_0) d\vec{r_0} dt_0.
 \label{cor}
\end{align}

We decompose the propagator $R$ into  $R_0$ and $R_1$, where $R_0$ is the equilibrium part, and, assuming that the perturbation is small, $R_1$ is the linear order perturbation:
\begin{equation}
	R \approx R_0+R_1
	\label{prpag}
\end{equation}
{Notice that we are considering linear response regime and therefore have neglected higher-order terms in the expansion.} This decomposition allows us to analyze the system's response to the perturbation in a systematic way, starting from the unperturbed equilibrium state $R_0$
and then considering the additional effects due to the perturbation captured by $R_1$. 
Using Eq.~\eqref{prpag}, we have:
\begin{equation*}
	\left[	 {\color{black}\left(\frac{\partial}{\partial t} + i \frac{2e}{\hbar c} \Phi(\vec{r}, t)\right)}  -D_{\varphi}\left(\hbar \nabla -2ie \vec{A}\right)^2 + \frac{1}{\tau_s} \right]\left(R_0+R_1\right)    =\delta(\vec{r}-\vec{r_0}) \delta(t-t_0).
\end{equation*}
which further gives:
\begin{equation*}	
	\left\lbrace
	\begin{aligned}
		&	\left[	 \frac{\partial}{\partial t}  -D_{\varphi}\left(\hbar \nabla -2ie \vec{A}\right)^2 + \frac{1}{\tau_s} \right]R_0    =\delta(\vec{r}-\vec{r_0}) \delta(t-t_0).\\
		& 	\left[	 \frac{\partial}{\partial t} +\left( -D_{\varphi}\left(\hbar \nabla -2ie \vec{A}\right)^2 + \frac{1}{\tau_s}\right)\right]R_1    =-  \left(i \frac{2e}{\hbar c} \Phi(\vec{r}, t)\right) R_0
	\end{aligned}
	\right.
\end{equation*}
Based on this equation, we derive the expression of $R_1$
   \begin{align}
	&R_1(\vec{r}, t; \vec{r'}, t') = -  \left(i \frac{2e}{\hbar c} \right)  \times \int R_0(\vec{r}, t; \vec{r}_1, t_1) R_0(\vec{r}_1, t_1; \vec{r'}, t') \Phi(\vec{r_1}, t_1) d\vec{r_1} dt_1
	\label{R1}
\end{align}  
Therefore, the total $R \approx R_0+R_1$ is constructed, where $R_0(\vec{r}, t; \vec{r'}, t')$ is given by Eq.~\eqref{R0} and $R_1(\vec{r}, t; \vec{r'}, t')$ is given by Eq.~\eqref{R1}. 
From the definition of the correlation function given by Eq.~\eqref{cor}, we have, 
\begin{equation}
C(\vec{r}, t; \vec{r}', t') \approx k_B T \int \left( R_0 + R_1 \right) \cdot \left( R_0^* + R_1^* \right) \, d\vec{r}_0 dt_0.
\end{equation}
Expanding this expression yields:
\begin{align}
C(\vec{r}, t; \vec{r}', t') &\simeq k_B T \int R_0(\vec{r}, t; \vec{r}_0, t_0) R_0^*(\vec{r}', t'; \vec{r}_0, t_0) \, d\vec{r}_0 dt_0 \nonumber \\
&\quad + k_B T \int \left[ R_0(\vec{r}, t; \vec{r}_0, t_0) R_1^*(\vec{r}', t'; \vec{r}_0, t_0) + R_1(\vec{r}, t; \vec{r}_0, t_0) R_0^*(\vec{r}', t'; \vec{r}_0, t_0) \right] \, d\vec{r}_0 dt_0 \nonumber \\
&\quad + \mathcal{O}(R_1^2).
\label{Cor}
\end{align}
The expansion of the correlation function can now be separated into its equilibrium and perturbative contributions. The leading-order term corresponds to the equilibrium state:
\begin{equation}
C_0(\vec{r}, t; \vec{r}', t') = k_B T \int R_0(\vec{r}, t; \vec{r}_0, t_0)\, R_0^*(\vec{r}', t'; \vec{r}_0, t_0) \, d\vec{r}_0 dt_0,
\end{equation}
while the first-order correction captures the system’s response to the external perturbation:
\begin{equation}
C_1(\vec{r}, t; \vec{r}', t') = k_B T \int \left[ R_0(\vec{r}, t; \vec{r}_0, t_0)\, R_1^*(\vec{r}', t'; \vec{r}_0, t_0) + R_1(\vec{r}, t; \vec{r}_0, t_0)\, R_0^*(\vec{r}', t'; \vec{r}_0, t_0) \right] d\vec{r}_0 dt_0.
\label{c1}
\end{equation}
Putting both contributions together, the total correlation function Eq.~\eqref{Cor} reads:
\begin{equation}
C(\vec{r}, t; \vec{r}', t') \simeq C_0(\vec{r}, t; \vec{r}', t') + C_1(\vec{r}, t; \vec{r}', t') + \mathcal{O}(R_1^2),
\end{equation}
where higher-order terms are neglected. The term \( C_1 \) governs the linear response of the system and serves as the starting point for constructing physical observables such as the heat current. It is important to clarify the methodology for studying thermoelectric transport. Instead of explicitly introducing a temperature gradient ($\nabla T$), which is challenging to implement consistently within the TDGL framework, we calculate the conjugate transport coefficient: the heat current generated in response to an applied electric field. This approach is standard in linear response theory, and the results are directly related to the Nernst effect via Onsager reciprocity relations \cite{Ussishkin2002PRL}.

{According to Eqs.~(22) and (23) of Ref.~\cite{Kazumi}, the heat-current operator $J^h$ is defined as}
\begin{equation}
J^h= -\hbar^2 D_\phi \left(\mathcal{A}+ \tilde{\mathcal{A}}\right)
C_1(\vec{r}, t; \vec{r}', t')\Big|_{\underset{\vec{r}=\vec{r}'}{t=t'}}
\label{heat}
\end{equation}
Here,
\begin{align*}
    &\mathcal{A}= \left(\nabla -\frac{2ie}{\hbar c} \vec{A}(\vec{r})\right)\left(\frac{\partial}{\partial t'} - i \frac{2e}{\hbar c} \Phi(\vec{r}')\right), \\
    &\tilde{\mathcal{A}}=\left(\nabla' +\frac{2ie}{\hbar c} \vec{A}(\vec{r}')\right)\left(\frac{\partial}{\partial t} + i \frac{2e}{\hbar c} \Phi(\vec{r})\right).
\end{align*}
 {\color{black}
At this stage it is important to distinguish the microscopic currents obtained from the correlator from the transport currents measured in an experiment. In a magnetic field one must separate transport and magnetization contributions according to~\cite{Ussishkin2002PRL}
\begin{equation}
\mathbf j^e = \mathbf j^e_{\mathrm{tr}} + \mathbf j^e_{\mathrm{mag}},
\qquad
\mathbf j^Q = \mathbf j^Q_{\mathrm{tr}} + \mathbf j^Q_{\mathrm{mag}},
\label{eq:current_split_final}
\end{equation}
with
\begin{equation}
\mathbf j^e_{\mathrm{mag}} = c\,\frac{\partial \mathbf M}{\partial T}\times(-\nabla T),
\qquad
\mathbf j^Q_{\mathrm{mag}} = c\,\mathbf M\times \mathbf E,
\label{eq:mag_currents_final}
\end{equation}
in Gaussian units, where $\mathbf M=M_z\hat z$ is the equilibrium orbital magnetization. The linear-response relations must therefore be written for transport currents,
\begin{equation}
\begin{pmatrix}
\mathbf j^e_{\mathrm{tr}}\\[2pt]
\mathbf j^Q_{\mathrm{tr}}
\end{pmatrix}
=
\begin{pmatrix}
\hat\sigma & \hat\alpha\\
T\hat{\tilde\alpha} & \hat\kappa
\end{pmatrix}
\begin{pmatrix}
\mathbf E\\[2pt]
-\nabla T
\end{pmatrix},
\qquad
\tilde\alpha_{ij}(B)=\alpha_{ji}(-B).
\label{eq:transport_matrix_final}
\end{equation}
For the geometry used below, $\mathbf E=E_y\hat y$ and $\mathbf B=B\hat z$, so that
\begin{equation}
j^{Q,\mathrm{tr}}_x
=
j^Q_x-(j^Q_{\mathrm{mag}})_x
=
j^Q_x + cM_zE_y.
\label{eq:transport_heat_current_final}
\end{equation}
Accordingly, the transverse thermoelectric transport coefficient is extracted from the transport heat current,
\begin{equation}
\alpha^{\mathrm{tr}}_{xy}=\frac{j^{Q,\mathrm{tr}}_x}{T E_y}.
\label{eq:alpha_transport_final}
\end{equation}
The experimentally measured Nernst signal under open-circuit conditions is then
\begin{equation}
\nu
\equiv
\frac{E_y}{(-\nabla_x T)B}
=
\frac{\alpha^{\mathrm{tr}}_{xy}\sigma_{xx}-\alpha^{\mathrm{tr}}_{xx}\sigma_{xy}}
{B\left(\sigma_{xx}^2+\sigma_{xy}^2\right)}.
\label{eq:nernst_transport_final}
\end{equation}
In the weak-field, particle-hole-symmetric limit one recovers the standard Gaussian-fluctuation scaling
\begin{equation}
\left|\alpha^{\mathrm{tr}}_{xy}\right|
\propto
\frac{\xi^2(T)}{\ell_B^2},
\qquad
\ell_B=\sqrt{\frac{\hbar}{2eB}},
\label{eq:alpha_xy_gaussian_final}
\end{equation}
for a two-dimensional layer.
}
 We consider an electric field along the $y$-direction, $\Phi(\vec{r}) = -yE_y$, and a magnetic field {\color{black}perpendicular to the $(x,y)$ plane},  represented by the vector potential $A_x = -By$. 
 
 The spatial differential operators $\nabla$ (and $\nabla'$), combined with $\vec{A}(\vec{r})$, ensure gauge invariance and capture the influence of the magnetic field on heat transport via the phase of the superconducting order parameter. The time derivatives $\partial_t$ (and $\partial_{t'}$), coupled with the scalar potential $\Phi(\vec{r})$, describe the system’s response to time-dependent electric fields, linking thermal and electrical effects.


{\color{black}
Therefore, the quantity obtained below should be understood as the microscopic/Kubo contribution to the transverse heat current. Comparison with the measurable Nernst response is made only after the magnetization-current subtraction in Eq.~\eqref{eq:transport_heat_current_final}. For completeness, spatial integration over a closed sample eliminates the net contribution of circulating magnetization currents through the sample boundaries, but the correct thermoelectric coefficient is still defined at the transport-current level as in Eqs.~\eqref{eq:transport_matrix_final}--\eqref{eq:alpha_transport_final}.
}

\noindent Substituting the expression of $C_1$ from Eq.~\eqref{c1} into the heat current formula Eq.~\eqref{heat}, we obtain:
\begin{align*}
	& C_1(\vec{r}, t; \vec{r}', t')\Big|_{\underset{\vec{r}=\vec{r}'}{t=t'}} \\
    &\hspace{0.2cm}=\left(\nabla -\frac{2ie}{\hbar c} \vec{A}(\vec{r})\right)\left(\frac{\partial}{\partial t'} - i \frac{2e}{\hbar c} \Phi(\vec{r}')\right) C_1(\vec{r}, t; \vec{r}', t')\Big|_{\underset{\vec{r}=\vec{r}'}{t=t'}}, \\
	&\hspace{0.2cm}=\left( \partial_x + \frac{2ie}{\hbar c} By\right)\left(\frac{\partial}{\partial t'} - i \frac{2e}{\hbar c} \Phi(\vec{r}')\right)\left(\frac{-i2e}{\hbar c}\right)
 \int \left(-E_y y_1\right) R_0(\vec{r}, t; \vec{r}_1, t_1)\, C_0^*(\vec{r}', t'; \vec{r}_1, t_1)\, dx_1\,dy_1\,dt_1.
\end{align*}
 The correlation function $C_0$ can be conveniently connected to the $R_0$ in Eq.~\eqref{R0} by the Fluctuation-Dissipation Theorem, and thus $C_0$ can be obtained from the imaginary part of $R_0$, and reads:

\begin{align}
	&C_0^*(\vec{r'}, t'; \vec{r_1}, t_1) \nonumber\\
    &\hspace{0.2cm}= \frac{k_B T}{2\pi}\int_{-\infty}^{\infty} e^{i \omega_2 (t_1-t')} d\omega_2. \int dk_{x_2} \sqrt{\frac{\omega_B}{2\pi \hbar D_\phi}}e^{i k_{x_2} (x_1-x')}  \times \exp\left(-\frac{(y'-y_{k_2})^2}{2a_B^2}\right)\exp\left(-\frac{(y_1-y_{k_2})^2}{2a_B^2}\right) \nonumber\\
	&\hspace{1.2cm} \times\sum_{n}\frac{1}{2^n. n!} \frac{H_n(\frac{(y'-y_{k_2})}{a_B}). H_n(\frac{y_1-y_{k_2}}{a_B})}{ \omega_2^2 + (\frac{1}{\tau_s}+(n+\frac{1}{2}) \hbar \omega_B)^2}   .
 \label{c0*t1}
\end{align}
To expand the expression for the heat current, we first substitute Eq.~\eqref{R0} and Eq.~\eqref{c0*t1} into Eq.~\eqref{c1} to obtain the full correlation function \( C_1(\vec{r}, t; \vec{r}', t') \). This result is then inserted into Eq.~\eqref{heat}, yielding the following analytical expression for the heat current:
\begin{align}
  	&J^h_{ x} =- k_B T E_y \frac{e}{c}\frac{\omega_B}{\pi}
    \int_{-\infty}^{+\infty} y_1\, dy_1
    \int_{-\infty}^{+\infty} dk_{x_1}
    \left(k_{x_1}+\frac{2eB}{\hbar c}y\right)
    \nonumber\\
    &\hspace{0.2cm}\times
    \exp\left(-\frac{(y-y_{k_{x_1}})^2}{a_B^2}-\frac{(y_1-y_{k_{x_1}})^2}{a_B^2}\right)
    \int_{-\infty}^{+\infty} \frac{d\omega_2}{2\pi}\, \omega_2^2
    \sum_{n=0}^{\infty}
    \frac{H_n(\frac{y-y_{k_{x_1}}}{a_B})\, H_n(\frac{y_1-y_{k_{x_1}}}{a_B})}
    {\omega_2^2 + \left(\frac{1}{\tau_s}+\left(n+\frac{1}{2}\right)\hbar\omega_B\right)^2}
    \nonumber\\
    &\hspace{1cm}\times
    \sum_{m=0}^{\infty}
    \frac{H_m(\frac{y-y_{k_{x_1}}}{a_B})\, H_m(\frac{y_1-y_{k_{x_1}}}{a_B})}
    {\omega_2^2 + \left(\frac{1}{\tau_s}+\left(m+\frac{1}{2}\right)\hbar\omega_B\right)^2}.
\label{heatex}
\end{align}
Eq.~\eqref{heatex} gives the analytical expression for the microscopic/Kubo contribution to the steady transverse heat current in the linear regime. It shows that the response is proportional to the applied electric field $E_y$, as expected in linear response. The exponential and Hermite-polynomial factors encode the spatial structure of the Landau-level modes; after spatial averaging, the resulting microscopic current becomes time independent in the steady state. This analytical result will be verified numerically in \figref{heat1} and \tabref{tab1}. It is important, however, to compare with experiment only at the transport-current level. In the weak-field, linear-response limit ($\omega_B\tau_s \ll 1$), after subtracting the magnetization-current contribution according to Eq.~\eqref{eq:transport_heat_current_final} and extracting the transport coefficient via Eq.~\eqref{eq:alpha_transport_final}, one recovers the standard Gaussian-fluctuation scaling
\begin{equation*}
\left|\alpha_{xy}^{\mathrm{tr}}\right| \propto \frac{\xi^2(T)}{\ell_B^2},
\qquad
\ell_B=\sqrt{\frac{\hbar}{2eB}},
\end{equation*}
for a two-dimensional layer, consistent with Ref.~\cite{Ussishkin2002PRL}.


\section{Numerical results of the linear type-I relaxation equation} \label{linsec}
 
In this section, we numerically solve the Type-I equation given by Eq.~\eqref{rdmf}. While the steady-state properties of this regime are established, our analysis focuses on the novel visualization and characterization of the spatiotemporal evolution of the propagator. This serves three critical purposes: first, it validates our numerical approach against analytical solutions; second, it provides a practical means to address the nonlinear Type-II relaxation equation in the next section; and third, it establishes a necessary baseline of linear diffusive dynamics. This baseline is essential to appreciate the dramatic, qualitative shift in transport mechanisms revealed in the Type-II regime. 

For the convenience of numerical simulation, we introduce the scaled variable \( \tilde{R} = \frac{1}{\tau_s} R \). Choosing the Landau gauge, \( \vec{A} = (-By, 0) \), Eq.~\eqref{rdmf} becomes:
\begin{align} \label{dimless}
\begin{split}
      \frac{\partial \tilde{R}}{\partial \tilde{t}}(\vec{\tilde{r}}, \tilde{t})
      - \tilde{D}_\varphi \left[ \frac{\partial^2}{\partial \tilde{y}^2}
      + \left( \frac{\partial }{\partial \tilde{x}}+i\tilde{y}\right)^2\right]\tilde{R}(\vec{\tilde{r}}, \tilde{t})
      + \tilde{R}(\vec{\tilde{r}}, \tilde{t})
      = \delta(\vec{\tilde{r}}-\vec{\tilde{r_0}})\delta(\tilde{t}-\tilde{t_0}).
\end{split}
\end{align}
where the dimensionless variables are defined as
\begin{equation*}
\frac{2eB}{\hbar}=l_B^{-2}, \qquad
\tilde{t}=\frac{t}{\tau_s}, \qquad
\tilde{\vec{r}}=\frac{\vec{r}}{l_B}, \qquad
\tilde{D}_\varphi=\frac{\tau_s D_\varphi \hbar^2}{l_B^2},
\end{equation*}
with $l_B=\sqrt{\hbar/(2eB)}$, and the dimensionless variables are defined as:
\begin{equation*}
\frac{eB}{\hbar} = l_B^{-2}, \qquad \tilde{t} = \frac{t}{\tau_s}, \qquad \tilde{\vec{r}} = \frac{\vec{r}}{l_B}, \qquad \tilde{D}_\varphi = \frac{\tau_s D_\varphi \hbar^2}{l_B^2}.
\end{equation*}
{\color{black} For clarity and reproducibility, we explicitly outline the normalization scheme used in the numerical simulations. Time is rescaled by the Ginzburg-Landau relaxation time $\tau_s$, such that $\tilde{t} = t/\tau_s$, while spatial coordinates are rescaled by the magnetic length $l_B = \sqrt{\hbar/(2eB)}$, i.e. $\tilde{\mathbf{r}} = \mathbf{r}/l_B$. The propagator is correspondingly rescaled as $\tilde{R} = R/\tau_s$, and the diffusion coefficient is expressed in dimensionless form as $\tilde{D}_\phi = \tau_s D_\phi \hbar^2 / l_B^2$. This normalization absorbs the fundamental physical scales into dimensionless parameters, leading to \eqref{dimless}, which is more suitable for numerical implementation. It also ensures that the results can be systematically reproduced and compared across different parameter regimes.}

We adopt the temporal gauge ($\Phi = 0$), so that the electromagnetic field is fully described by the vector potential $\vec{A} = (-By, 0)$. 
Eq. \eqref{dimless} is then solved numerically using the $\delta_{-}$Ziti method, with the analytical reference solution given by Eq.~\eqref{R0}.  Observe that, The propagator exhibits exponential decay in time. Indeed, For \( n = 0 \), the denominator in the summation reduces to  $i \omega_1 + \frac{1}{\tau_s} + \frac{1}{2} \hbar \omega_B$, and the integral over \( \omega_1 \) contains the factor \( e^{i \omega_1 (t - t_1)} \). This leads to a decay governed by the imaginary part of the pole at  $\omega_1 = -i\left(\frac{1}{\tau_s} + \frac{1}{2} \hbar \omega_B\right). $ Since \( \hbar \omega_B \) introduces only a constant frequency shift, the dominant time dependence is given by $\exp\left(-\frac{t}{\tau_s}\right).$ 

\begin{figure}[H] 
\centering
 \begin{tikzpicture}
    \node[anchor=south west, inner sep=0] (img) at (0,0) 
        {\includegraphics[height=2.5in,width=8.5cm]{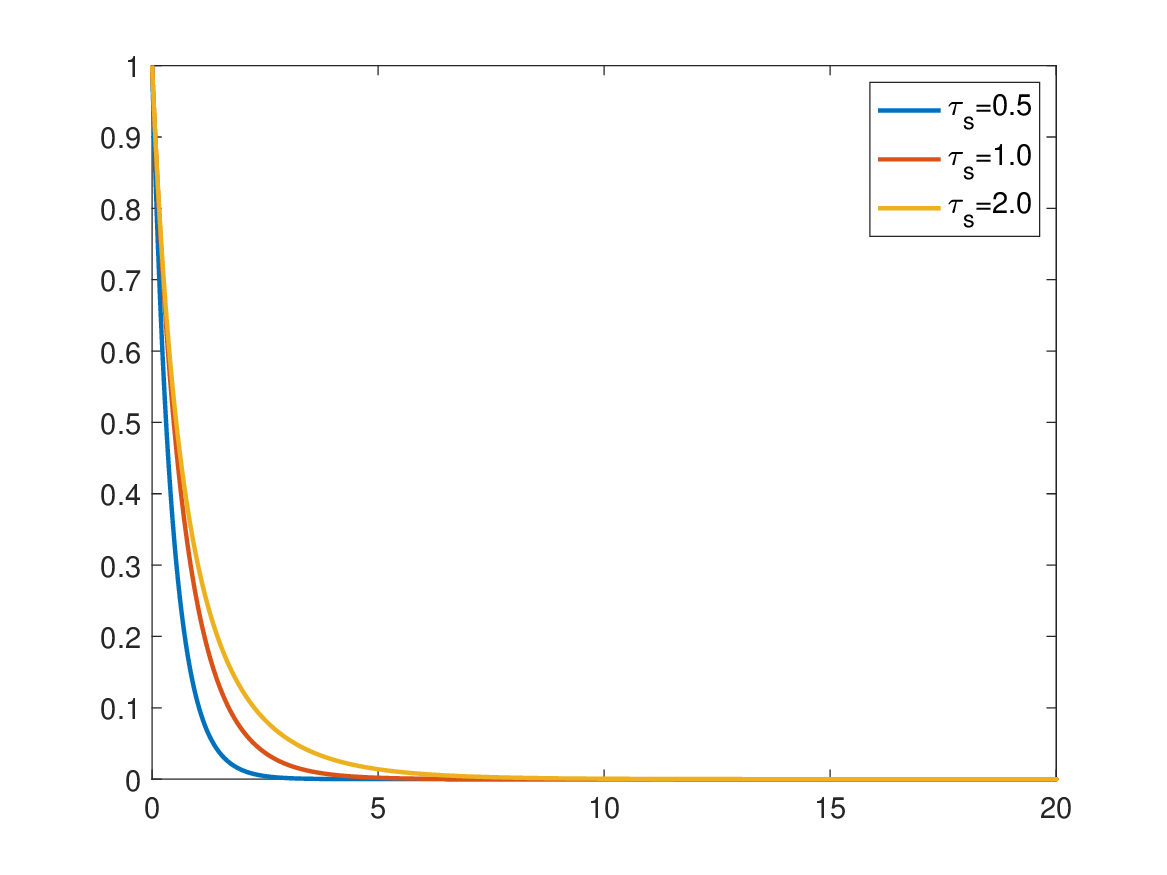}};
 \node[rotate=90] at (0 ,2.8) {{Evolution of the peak}};
        \node at (4,0) {{Time}};

\end{tikzpicture}  
\caption{\justifying {Time evolution of the real part of the numerical propagator from $t=0$ to $t=20$, for several values of the relaxation time $\tau_s$. The dimensionless equation was solved, with time in the x-axis given by $t/\tau_s$. The solution exhibits exponential decay, governed by the linear relaxation term $\frac{1}{\tau_s} R_0$, without any saturation due to the absence of nonlinear effects. Smaller values of $\tau_s$ lead to faster decay, while larger values of $\tau_s$ result in slower decay. These results illustrate the role of $\tau_s$ in controlling the decay rate of the solution, providing insights into the relaxation dynamics of systems in the Type-I relaxation regime near the critical temperature $T_c$.}}
    \label{decay}
\end{figure}

To accurately reproduce this behavior, we numerically solve Eq.~\eqref{dimless} using the $\delta_{-}$Ziti method. This approach is particularly well-suited for handling relaxation-type dynamics, allowing for precise treatment of both spatial diffusion and temporal decay. The essential aspects of this numerical method are summarized in Appendix~\ref{numapp}, and we refer the interested reader to \cite{rajeq,Aa,Ks} for a more comprehensive discussion. The numerical scheme is constructed to balance all terms in the discretized equations. Stability is ensured through careful selection of the time step and spatial discretization parameters, which helps prevent artificial numerical dissipation and preserves the system's intrinsic decay dynamics. \figref{decay} shows the time evolution of the numerical solution for various values of $\tau_s$, clearly demonstrating how the relaxation time controls the rate of decay. The strong agreement between numerical and analytical results confirms the robustness of the method for modeling relaxation-driven diffusive processes. 
{\color{black}
All the numerical simulations are performed on a finite computational domain large enough that, over the early-time window shown in the main figures, the evolving response remains localized away from the boundaries. In practice, the calculations reported in the manuscript are carried out on the domain $[-1,1]\times[-1,1]$ using a uniform grid of $201\times 201$ points and a dimensionless time step $\Delta\tilde t=10^{-3}$. Convergence was checked by halving $\Delta\tilde t$ and doubling the spatial resolution. All quantities are expressed in the dimensionless variables defined in Eq.~\eqref{dimless}. The purpose of the square-versus-circular comparison in Appendix~B is therefore to test whether the early-time four-lobe structure is imposed by the square box geometry. In this sense, Appendix~B supports the claim that the early-time pattern is not a trivial artifact of the square domain, while the late-time redistribution is naturally more sensitive to the finite computational boundary.}

After following  the time evolution of $R_0$, we now examine the spatial profile  of the numerical propagator at different time. Using the analytical solution at $t=0$ as an initial condition, we aim to observe how the solution evolves in both amplitude and spatial distribution, illustrating the combined effect of relaxation and diffusion.

\begin{figure}[H] 
\centering 
  \begin{tikzpicture}
    \node[anchor=south west, inner sep=0] (img) at (0,0) 
        {\includegraphics[height=2.5in,width=8.5cm]{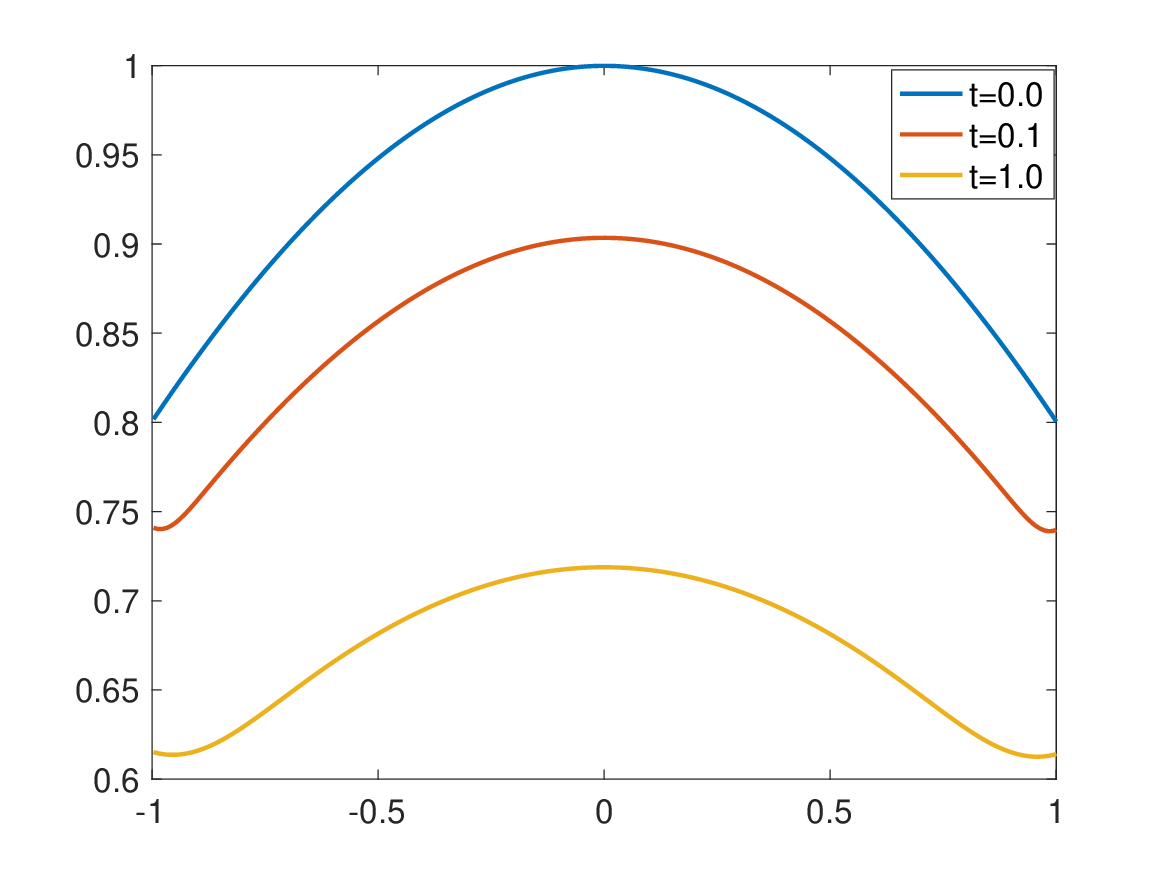}};
 \node[rotate=90] at (0 ,2.8) {{Space distribution}};
    \node at (4,0) {{$x-$ axis}};
\end{tikzpicture}
		\caption{\justifying One dimensional space distribution of the numerical propagator, at several times. The initial condition is derived from the analytical solution at $t=0$. At small times, the solution decays in amplitude due to the linear relaxation term $\frac{1}{\tau_s} R_0$, while simultaneously broadening in space due to the diffusion term $D_{\varphi}(\hbar \nabla-2 i e \vec{A})^2$. As time progresses, the solution continues to decay and broaden, reflecting the combined effects of relaxation and diffusion. These results illustrate the role of relaxation and diffusion in shaping the spatial and temporal evolution of the propagator in the Type-I relaxation regime.}
	\label{sdtype1}
\end{figure} 

\figref{sdtype1} shows the gradual decay and spatial broadening of the propagator over time. The decrease in amplitude reflects the effect of relaxation, while the widening of the distribution is governed by diffusion, consistent with the theoretical expectations.

\begin{figure}[H] 
    \centering 
    \begin{tikzpicture}
        \node[inner sep=0pt] (A) %
         {\includegraphics[height=2.5in,width=8.5cm]{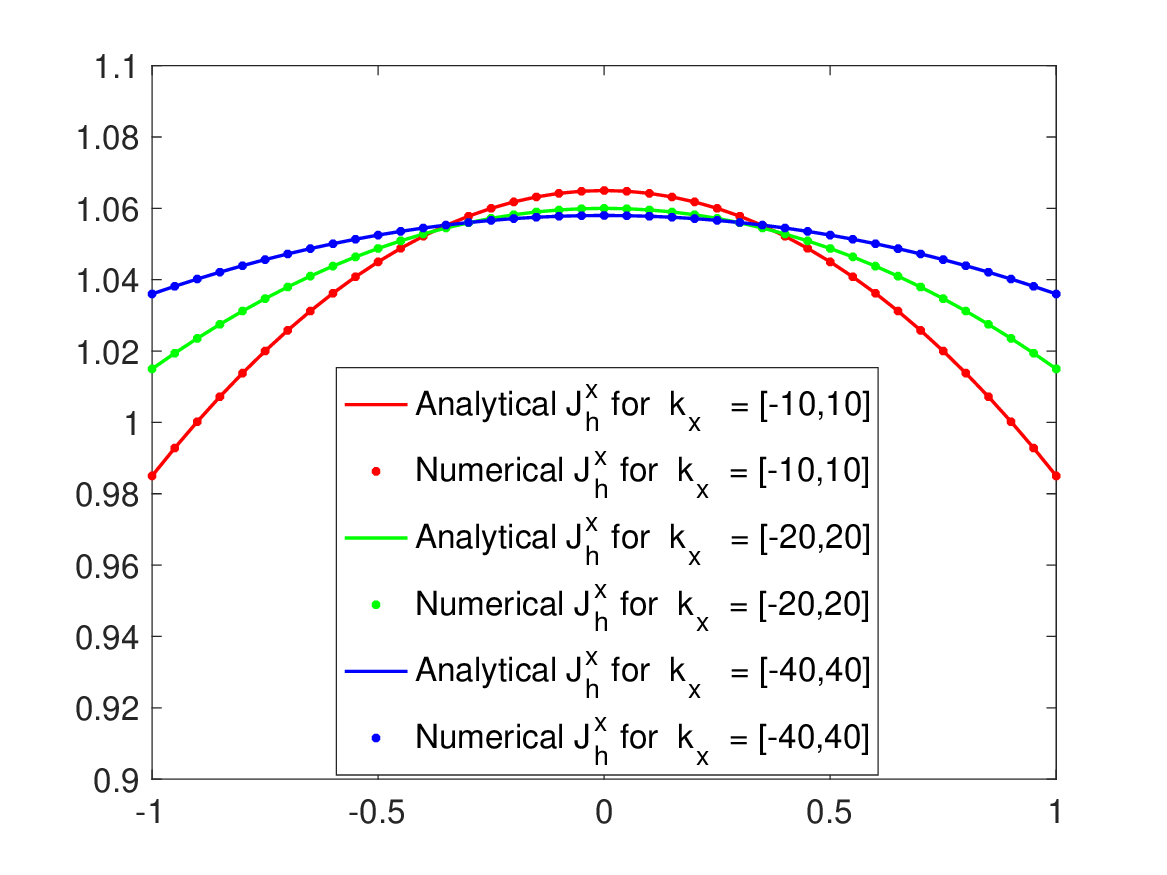}};
        \node[rotate=90] at (-4 ,0.3) {{Analytical and numerical} $J^h_x$};
        \node at (0,-2.9) {{$y-$ axis}};
        \label{heatap}
    \end{tikzpicture}
    \caption{\justifying Comparison of the analytical heat current (continuous line) with the approximated heat current computed using the $\delta_{-}$Ziti method (dotted line). The analytical solution is derived from the exact propagator of the system, while the $\delta_{-}$Ziti method provides a numerical approximation of the heat current.  }
    \label{heat1}
\end{figure}

It is worth pointing out that the analytical expression of the heat current in Eq.~\eqref{heatex} depends on the spatial coordinate $y$, although for a sufficiently large system, the current is expected to be uniform along $ y $. To investigate this, we compute the heat current for a range of $k_x\in [-40,40]$ and compare the spatial averages of the analytical and numerical results. As shown in \figref{heat1}, the numerical heat current, despite being derived from a time-space dependent model, remains stable over time and aligns closely with the analytical results. \vskip 2pt

The average values, reported in \tabref{tab1}, exhibit good agreement with relative errors on the order of  $10^{-4}$, validating the consistency of the numerical approach. \vskip 2pt 
{\renewcommand{\arraystretch}{1.5}
	\begin{table}[h!]
		\begin{center}
			\begin{tabular}{  p{2cm}   p{2cm}   p{2cm}   p{2cm}  }
				\hline \hline 
				$k_x$&$[-10, 10]$&$[-20, 20]$&$[-40, 40]$\\
				\hline \hline 
				\multicolumn{4}{c}{$t=1$}\\
				\hline \hline 
				Average$_{ex}$ &$1.035$ &$  1.0496$&	$1.0521$ \\ 
				Average$_{ap}$ &$1.0351$ &	$  1.0495$&	$1.0521$ \\
				\hline \hline 
						\multicolumn{4}{c}{$t=2$}\\
				\hline \hline 
			Average$_{ex}$ &$1.0351$ &$  1.0498$&	$1.0521$ \\ 
				Average$_{ap}$ &$1.0352$ &	$  1.05$&	$1.0522$ \\
				\hline \hline 
				
						\multicolumn{4}{c}{$t=20$}\\
				\hline \hline 
		Average$_{ex}$ &$1.0351$ &$  1.0498$&	$1.0521$ \\ 
				Average$_{ap}$ &$1.0352$ &	$  1.05$&	$1.0522$ \\
				\hline \hline 
				
			\end{tabular}
		\end{center}
		    \caption{\justifying  Comparison of exact and numerical heat current averages across different $k_x$ domains at varying time instances. The exact averages (denoted as Average$_{e x}$ ) are computed from the analytical solution, while the numerical averages (denoted as Average$_{a p}$ ) are obtained using the $\delta_{-}$Ziti method. The table shows the results for $k_x$ domains $[-10,10],[-20,20]$, and $[-40,40]$ at times $t=1, t=2$, and $t=20$.}
		    		\label{tab1}
	\end{table}
}

The close agreement between the numerical and analytical results strongly supports the accuracy and stability of the used  method in capturing heat transport phenomena. This consistency highlights the method's reliability for systems with known solutions and demonstrates its potential for addressing more complex scenarios, such as the Type-II relaxation equation, where analytical solutions are unavailable. In the next section, we present our  numerical findings for Type-II systems, exploring their dynamics and heat flow behavior. Based on the validated framework from the Type-I case, we aim to gain deeper insight into the interplay between diffusion, relaxation, and nonlinearity in the Type-II relaxation regime below $T_c$, shedding light on the fundamental mechanisms governing the resulting transient dynamics.

\begin{figure}[!htbp] 
    \centering
    \captionsetup{justification=raggedright, singlelinecheck=false}
    \begin{tikzpicture}
        \node[inner sep=0pt] (A) {\includegraphics[height=2.5in,width=9cm]{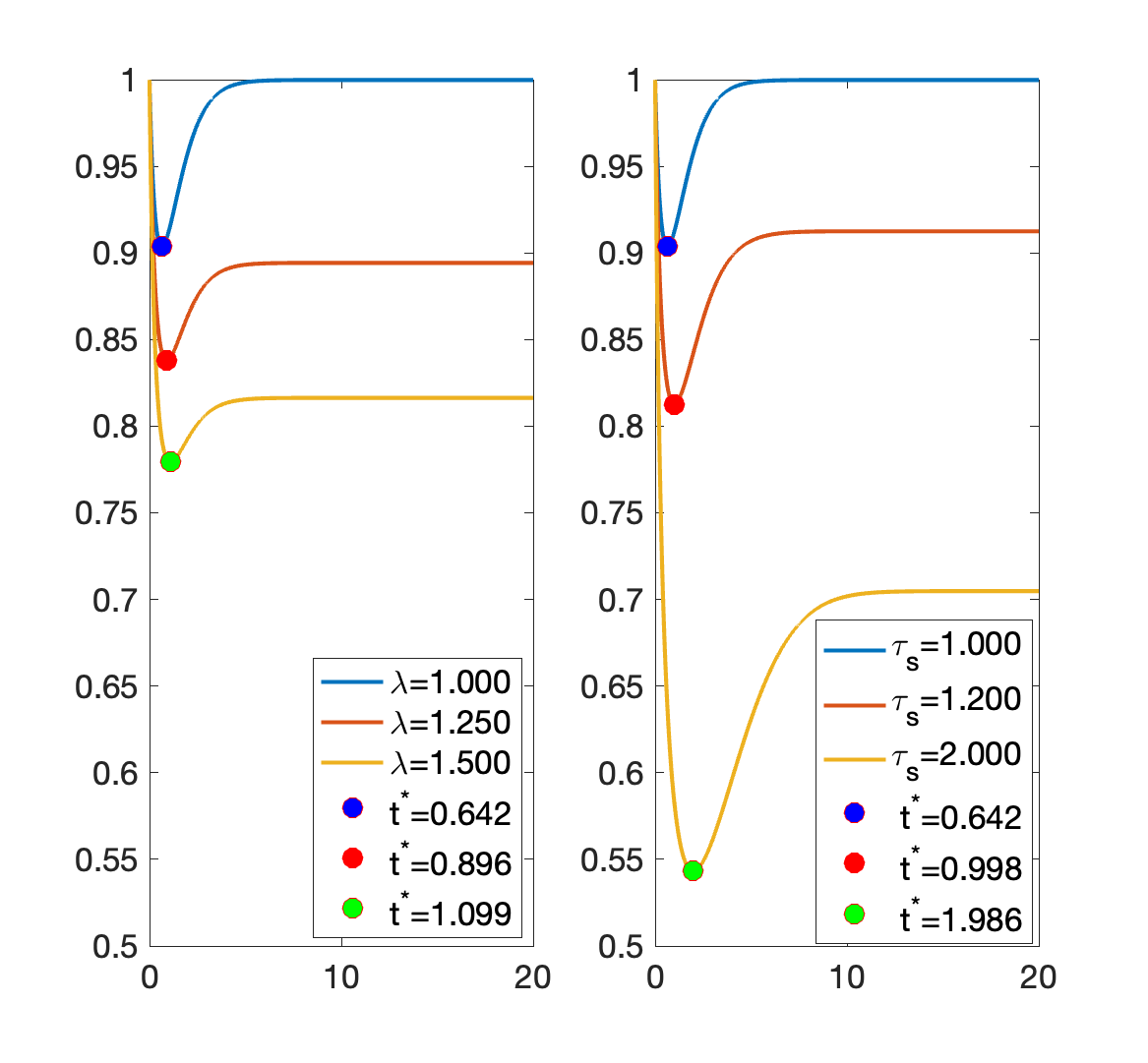}};
        \node[rotate=90] at (-4.25, 0.4) {{Evolution of the peak}};
        \node at (0, -3.3) {{Time}};
    \end{tikzpicture}
    \caption{{\color{black}
Time evolution of the peak of the solution $R_0(t)$ from $t=0$ to $t=20$. \textbf{Left panel:} The relaxation time is fixed at $\tau_s=1$, and the nonlinearity parameter $\lambda$ is varied $(\lambda=1,\lambda=1.25,\lambda=1.5)$. The solution exhibits an initial decay at small times, followed by a transition toward saturation at larger times. The transition time $t^*$ depends on $\lambda$: for $\lambda=1$, $t^*=0.642$; for $\lambda=1.25$, $t^*=0.896$; and for $\lambda=1.5$, $t^*=1.099$. The saturation value is inversely proportional to $\sqrt{\lambda\tau_s}$, reflecting the balance between the linear term $\frac{1}{\tau_s}R_0$ and the nonlinear term $\lambda R_0^3$. \textbf{Right panel:} The nonlinearity parameter is fixed at $\lambda=1$, and the relaxation time $\tau_s$ is varied $(\tau_s=1,\tau_s=1.2,\tau_s=2)$. Smaller values of $\tau_s$ lead to faster decay and earlier saturation, while larger values delay the transition. Specifically, for $\tau_s=1$, $t^*=0.642$; for $\tau_s=1.2$, $t^*=0.998$; and for $\tau_s=2$, $t^*=1.986$. These results highlight the role of $\tau_s$ and $\lambda$ in controlling the decay rate, transition time, and saturation value of the solution.
}}
    \label{peakt2}
\end{figure}

\section{Numerical results of the nonlinear Type-II relaxation equation}\label{nonlinsec}

{\color{black}
Below $T_c$, the deterministic part of the full time-dependent Ginzburg--Landau (TDGL) equation may be written as~\cite{Schmid1966,LarkinVarlamov2005,Rosenstein2010}
\begin{equation}
\Gamma^{-1}
\left(
\partial_t+\frac{i 2e}{\hbar}\Phi
\right)\psi
=
-a(T,B)\psi
-b|\psi|^2\psi
+K\left(\nabla-\frac{i 2 e}{\hbar c}\vec{A}\right)^2\psi,
\label{eq:full_tdgl_below_Tc}
\end{equation}
with $a(T,B)<0$ below the transition and $b>0$. The quartic GL free-energy term therefore generates the cubic TDGL nonlinearity $-b|\psi|^2\psi$, which is the leading local mechanism that saturates the growth of the order-parameter magnitude below $T_c$. Writing $\psi=\rho e^{i\theta}$ and taking the real part of Eq.~\eqref{eq:full_tdgl_below_Tc} gives, schematically,
\begin{equation}
\Gamma^{-1}\partial_t \rho
=
-a(T,B)\rho
-b\rho^3
+K\left[
\nabla^2\rho
-\rho\left(\nabla\theta-\frac{2e}{\hbar c}\vec{A}\right)^2
\right].
\label{eq:amplitude_sector}
\end{equation}
Equation~\eqref{eq:amplitude_sector} highlights the two ingredients relevant to the present work: the linear instability below $T_c$ and its nonlinear saturation by the cubic term. It also makes clear that the full amplitude dynamics remain coupled to phase gradients and hence to vortex physics.

In the present paper we do not attempt to solve the full complex mixed-state TDGL problem. Instead, to isolate the short-time local recovery after a localized perturbation, we introduce a reduced nonlinear response field, still denoted by $R_0(\vec{r},t;\vec{r}_0,t_0)$, and study the fixed-gauge nonlinear evolution equation
\begin{align}
\left[
\frac{\partial}{\partial t}
- D_{\varphi}\left(\hbar \nabla - 2ie \vec{A}\right)^2
-\frac{1}{\tau_s}
\right]
R_0(\vec{r},t;\vec{r}_0,t_0)
+\lambda R_0^3(\vec{r},t;\vec{r}_0,t_0)
=
\delta(\vec{r}-\vec{r}_0)\delta(t-t_0),
\label{type2}
\end{align}
where the coefficients are interpreted as effective coarse-grained parameters, $\frac{1}{\tau_s}\sim \Gamma |a_{\mathrm{eff}}(T,B)|, \lambda\sim \Gamma b$, with $a_{\mathrm{eff}}(T,B)$ measuring the local distance to the instability line after coarse graining over orbital and slowly varying phase effects. The corresponding homogeneous saturation scale, $R_{\mathrm{sat}}=\frac{1}{\sqrt{\lambda\tau_s}}$, is the reduced-model analogue of the GL amplitude $|\psi|_{\mathrm{eq}}=\sqrt{|a_{\mathrm{eff}}|/b}$.

We emphasize that Eq.~\eqref{type2} is not the reduced retarded response of the full TDGL theory. Strictly speaking, the exact retarded kernel of a nonlinear theory is a functional derivative around a chosen background configuration. The field $R_0(\vec{r},t;\vec{r}_0,t_0)$ used here is therefore interpreted as a reduced retarded response envelope or reduced nonlinear response field obtained from a causal initial-value problem in a fixed Landau gauge, appropriate to the early-time amplitude-controlled regime after a localized perturbation. It is not intended as a complete theory of long-time vortex thermoelectric transport. Correspondingly, the current patterns extracted from this equation should be read as effective local heat-current textures within the same reduced model, not as exact Kubo transport coefficients of the full mixed state. The reduced model in Eq.~\eqref{type2} retains diffusion, the local linear instability below $T_c$, and the leading local saturation term. It does not include an explicit evolution equation for the phase field $\theta$ and therefore does not describe phase slips, vortex nucleation and annihilation, vortex viscosity and flux flow, Hall-like vortex drift, or entropy carried by mobile vortices~\cite{CaroliMaki1967,Podolsky2007,BehniaAubin2016}. The Type-II results reported below should therefore be interpreted as local transient signatures of nonlinear amplitude-controlled relaxation, not as a complete theory of mixed-state vortex thermoelectric transport.} 

{\color{black}
For the nonlinear Type-II simulations it is useful to normalize the response amplitude by its homogeneous saturation value,
\begin{equation}
R_*=(\lambda\tau_s)^{-1/2},
\qquad
\tilde R = \frac{R_0}{R_*},
\end{equation}
together with
\begin{equation}
\tilde t=\frac{t}{\tau_s},
\qquad
\tilde{\vec r}=\frac{\vec r}{\ell_B},
\qquad
\tilde D_\varphi = \frac{\tau_s D_\varphi \hbar^2}{\ell_B^2},
\qquad
\ell_B=\sqrt{\frac{\hbar}{2eB}}.
\end{equation}
Equation~\eqref{type2} then becomes
\begin{equation}
\partial_{\tilde t}\tilde R
-
\tilde D_\varphi
\left[
\partial_{\tilde y}^2
+
\left(\partial_{\tilde x}+i\tilde y\right)^2
\right]\tilde R
-\tilde R+\tilde R^3
=
\delta(\tilde{\vec r}-\tilde{\vec r}_0)\delta(\tilde t-\tilde t_0).
\label{eq:typeII_dimensionless_final}
\end{equation}
This normalization makes explicit that the spatially uniform long-time solution saturates at $\tilde R=1$, i.e.\ $R_0=R_*=(\lambda\tau_s)^{-1/2}$.
}

Due to the complexity introduced by the non-linear term $R_0^3$ and the damping contribution, an analytical solution to this equation is not feasible. Instead, we employ the $\delta_{-}$Ziti method  to compute the time evolution of the propagator, as shown in \figref{peakt2}. The results reveal a rich interplay between diffusion, relaxation, and nonlinear effects. Initially, at early times \( (t \ll \tau_s) \), the solution exhibits an initial decay dominated by the diffusion term, consistent with the diffusive nature of the system. In this regime, the nonlinear contribution \( \lambda R_0^3 \) remains negligible, and the dynamics follow a classical diffusion process. However, as time progresses, nonlinear effects accumulate: the cubic term $\lambda R_0^3$ becomes significant enough to inhibit further decay, effectively stabilizing the solution at a saturation value, inversely proportional to $\sqrt{\lambda \tau_s}$. The non-monotonic evolution of the peak response amplitude stands in sharp contrast to the exponential decay found in the Type-I regime.

\begin{figure}[!htbp] 
    \centering
    \captionsetup{justification=raggedright, singlelinecheck=false}
    \begin{tikzpicture}
        \node[inner sep=0pt] (A) {\includegraphics[height=2.5in,width=8.5cm]{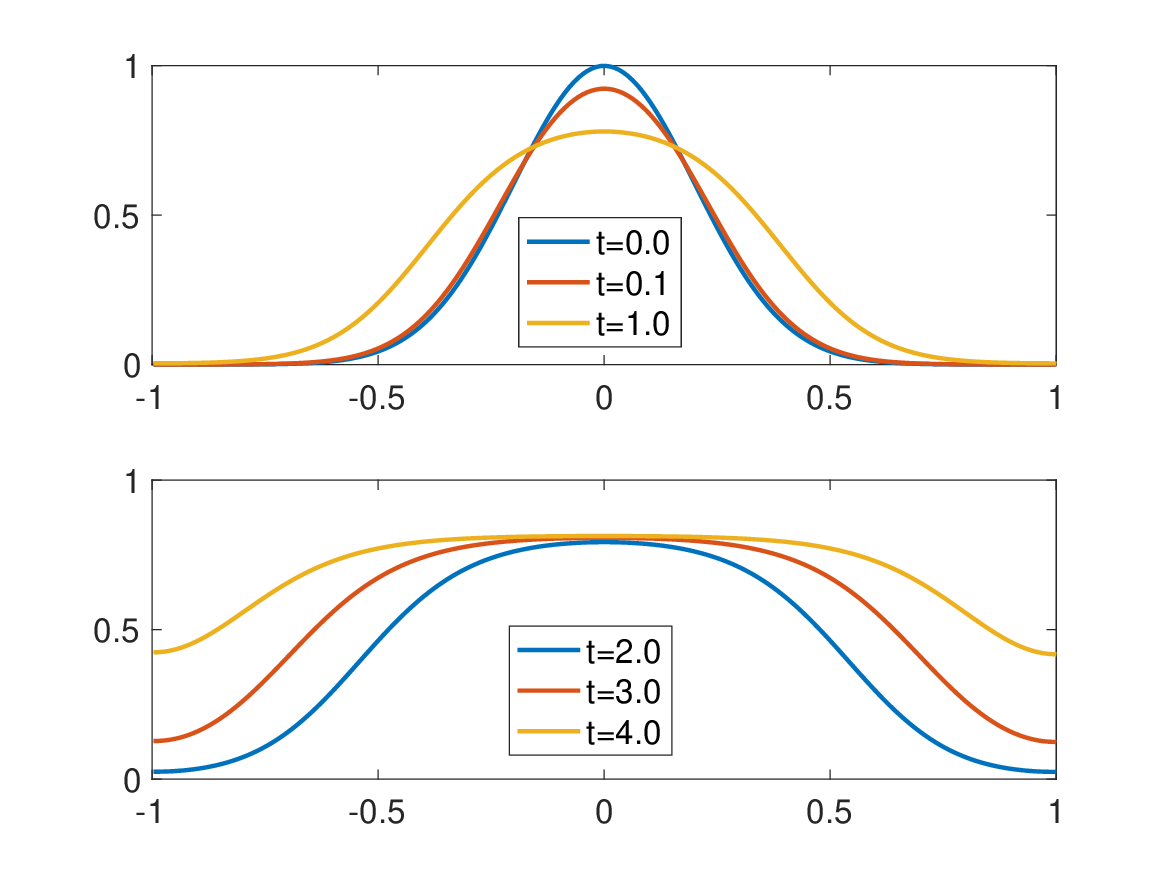}};
        \node[rotate=90] at (-3.8 ,0.3) {{Space distribution}};
        \node at (0,-3) {{$x-$ axis}};
    \end{tikzpicture}
    \caption{Spatial distribution of the solution $R_0(x,t)$ at different time, illustrating the evolution from early to later stages. The initial condition is a Gaussian centered at $x=0$ with peak value $1$. At early times, the central peak decreases mainly because diffusion redistributes the initially localized response, while the profile broadens in space. At later times, the dynamics reflect the competition among diffusion, the below-$T_c$ linear growth tendency, and the nonlinear saturation term $\lambda R_0^3$. The spatial broadening highlights the diffusive character of the dynamics, whereas the bounded long-time amplitude reflects nonlinear saturation.}
    \label{sdtype2}
\end{figure}
The spatial distribution of the propagator for the Type-II equation is plotted in \figref{sdtype2}, providing further insight into the interplay among diffusion, the below-$T_c$ linear growth tendency, and nonlinear saturation. This illustration, captured at different time steps, highlights how the spatial profile evolves under this competition. At early times, the initially localized peak at the center decreases because diffusion rapidly redistributes the concentrated response, while the profile spreads outward. As time progresses, the cubic term becomes increasingly important and limits further growth, so the dynamics cross over to a broadened but bounded profile. We therefore interpret the early-time decrease of the central peak as diffusion-dominated redistribution rather than as relaxation induced by the below-$T_c$ linear term itself. \vskip 2pt

\begin{figure*}[!htbp] 
    \centering  
    \minipage{0.33\textwidth}
        \includegraphics[height=1.4in, width=1\linewidth]{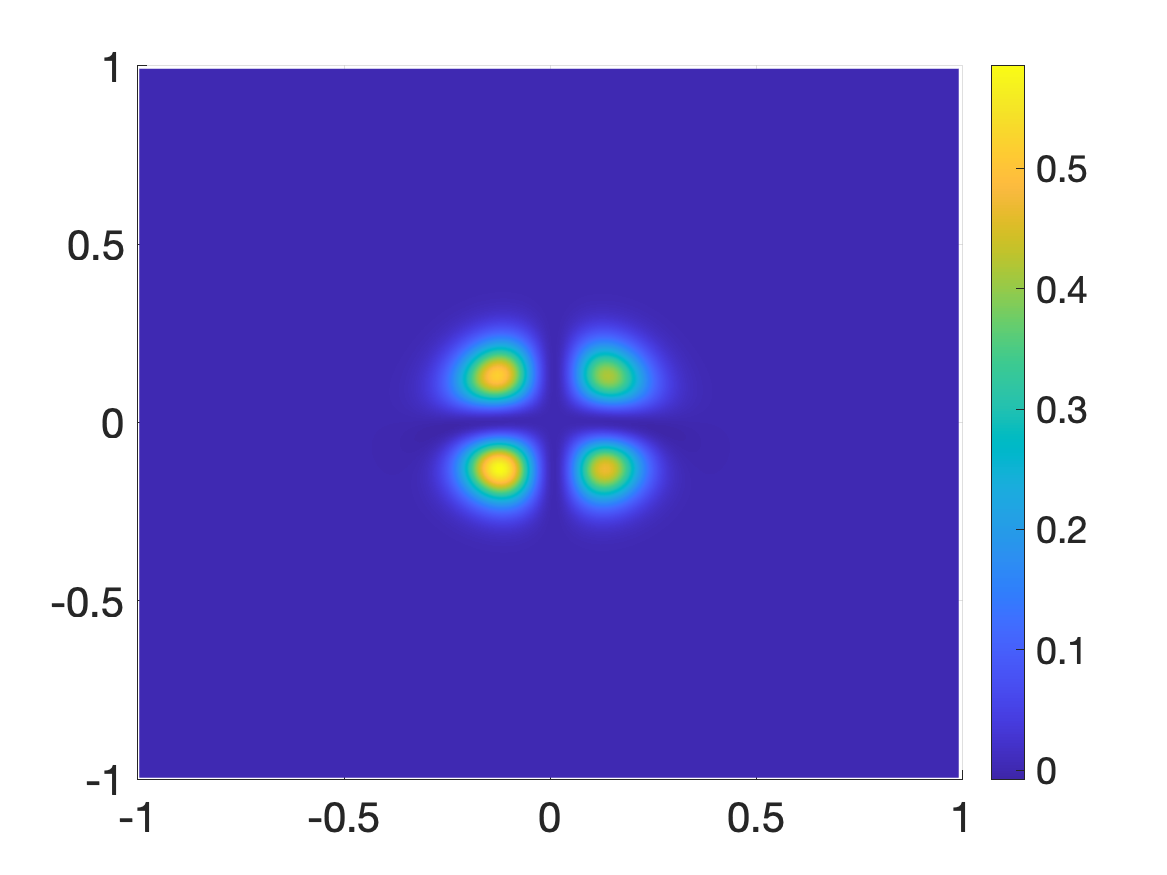}
        \subcaption{$t=0.1$}
    \endminipage
    \minipage{0.33\textwidth}
        \includegraphics[height=1.4in, width=2.3in]{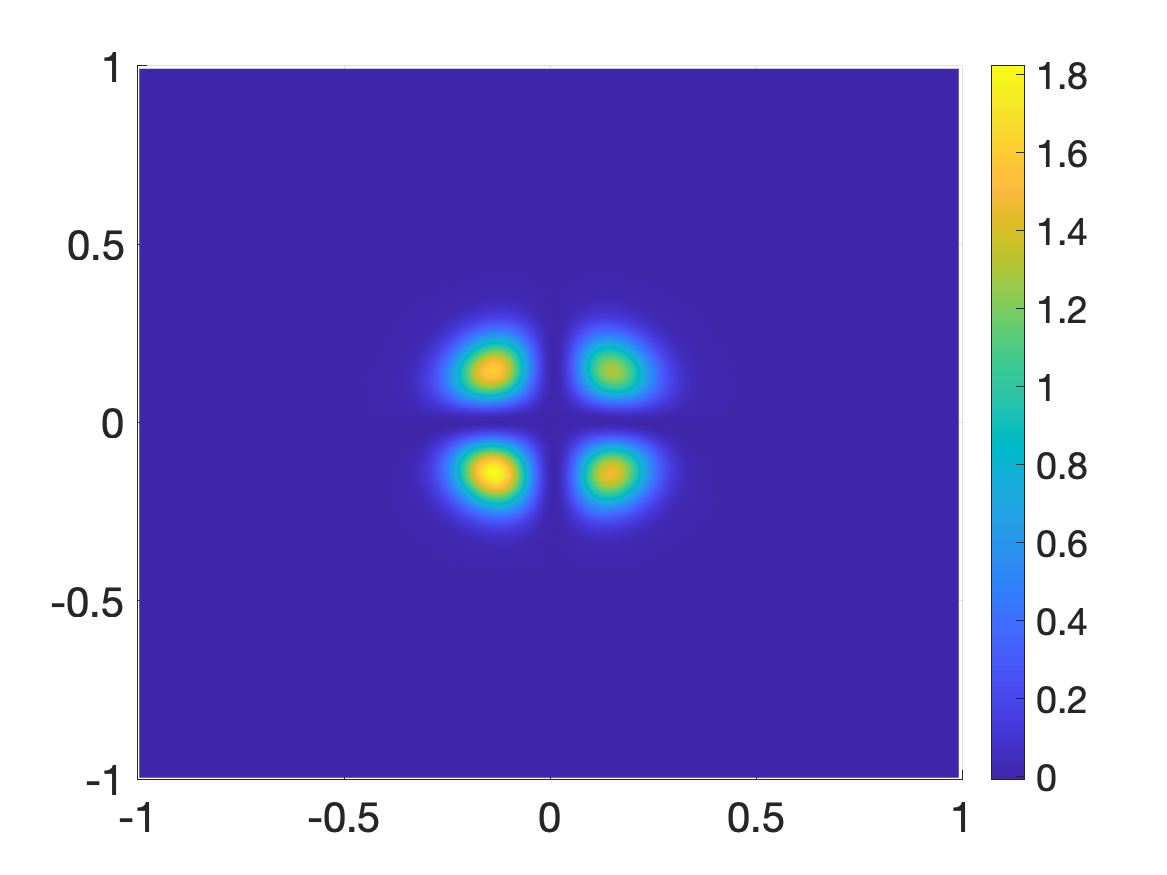}
        \subcaption{$t=0.2$}
    \endminipage
    \minipage{0.33\textwidth}
        \includegraphics[height=1.4in, width=2.3in]{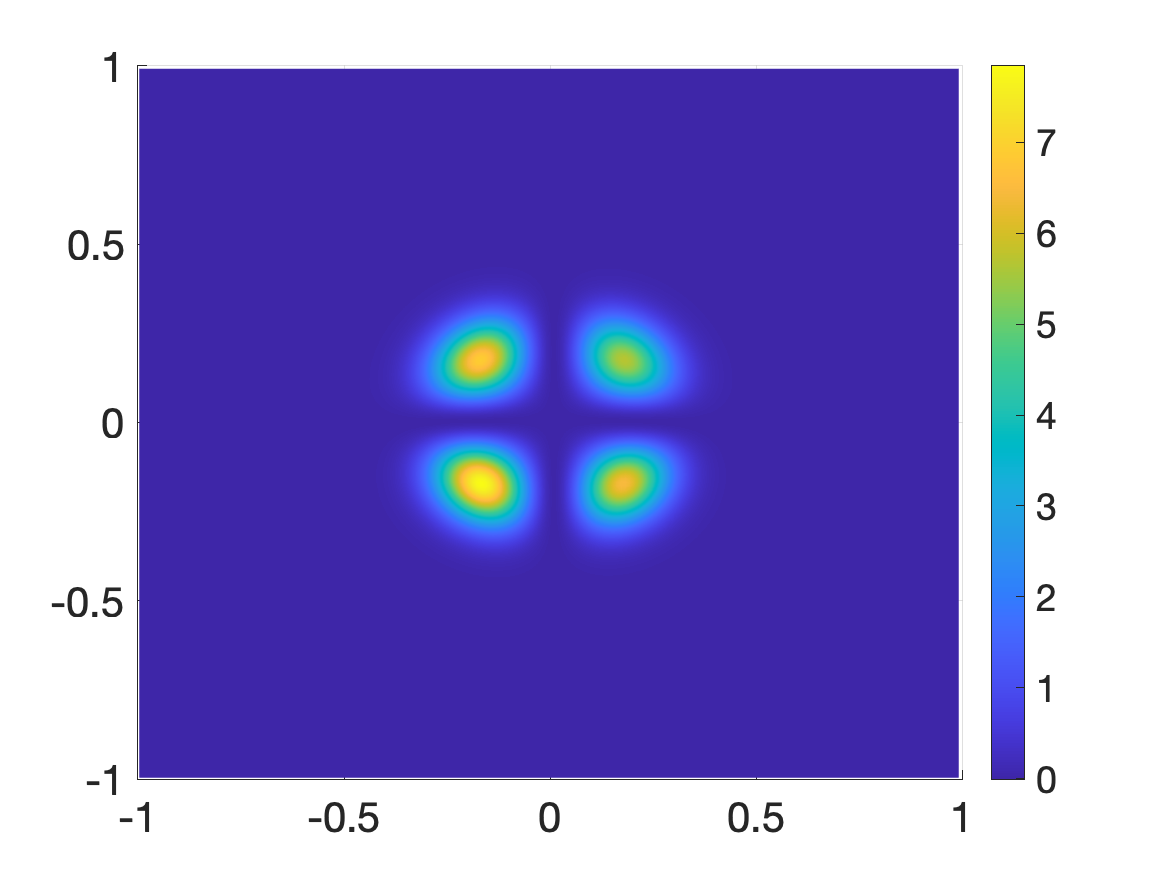}
        \subcaption{$t=0.5$}
    \endminipage
    \\ 
    \minipage{0.33\textwidth}
        \includegraphics[height=1.4in, width=2.3in]{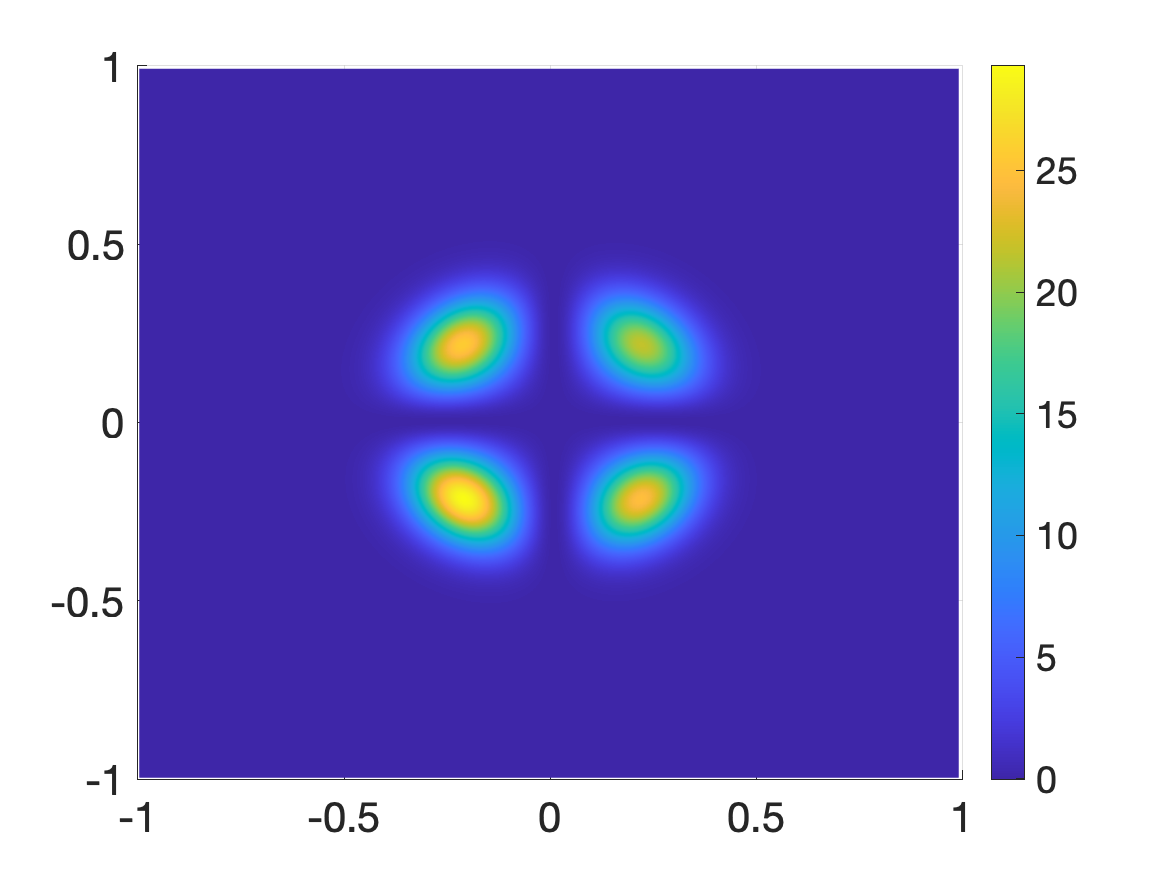}
        \subcaption{$t=1.0$}
    \endminipage
    \minipage{0.33\textwidth}
        \includegraphics[height=1.4in, width=2.3in]{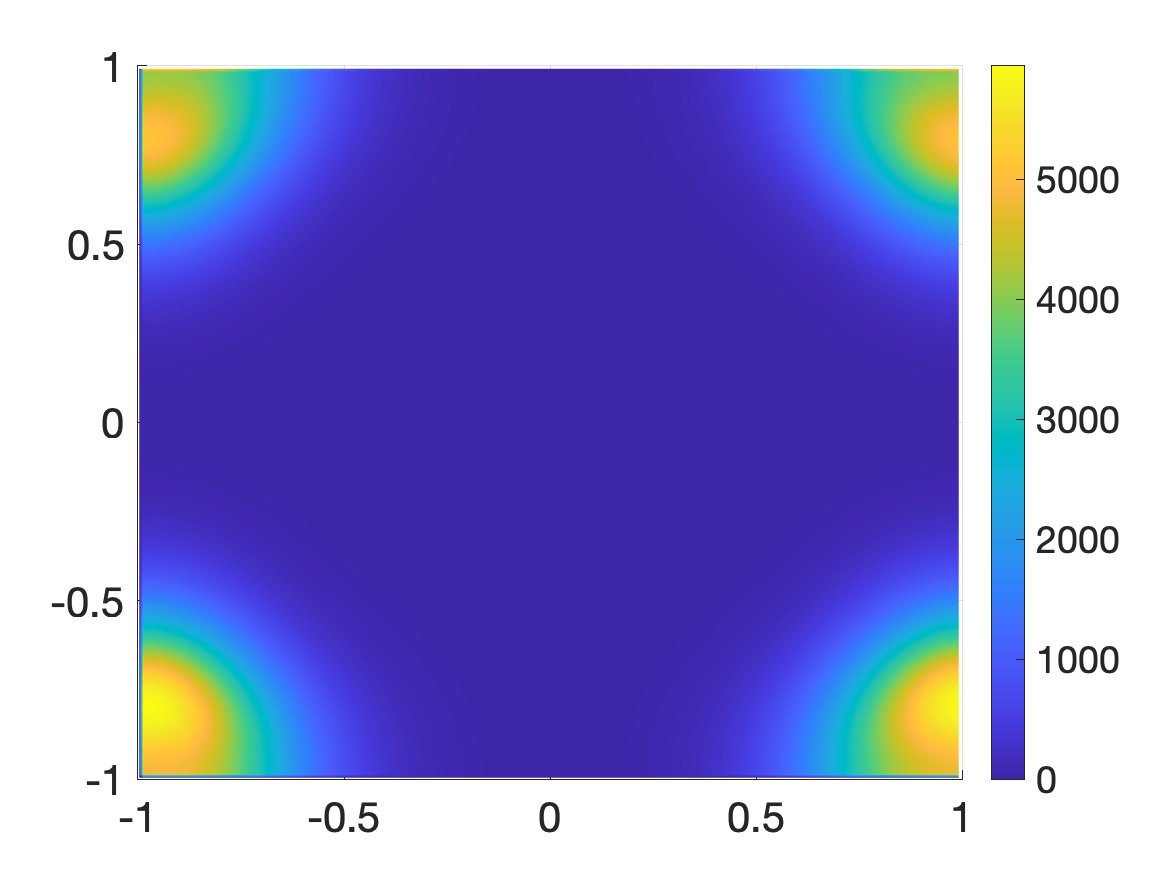}
        \subcaption{$t=6.0$}
    \endminipage
    \minipage{0.33\textwidth}
        \includegraphics[height=1.4in, width=2.3in]{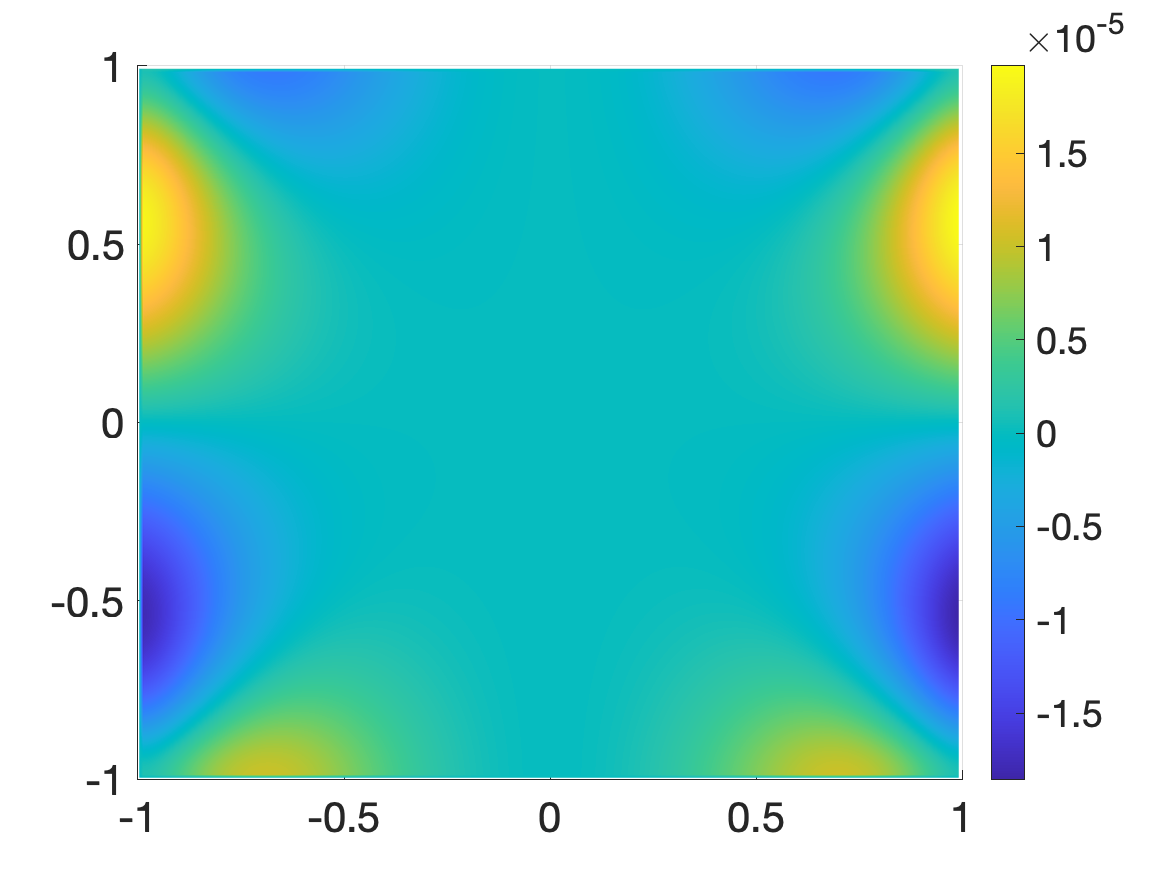}
        \subcaption{$t=12.0$}
    \endminipage
    \caption{\justifying {The numerical heat current at different time. The figure shows  the numerical heat current at various time steps, labeled (a) through (f). (a) $t=0.1$ : A quadrupole structure appears at the center with four distinct high-intensity regions. (b) $t=0.2$ : The peak intensity increases while the quadrupole structure remains localized. (c) $t=0.5$ : The heat current continues to grow in magnitude while maintaining the four-region pattern at the center. (d) $t=1$ : The intensity further increases, and the heat begins to spread outward. (e) $t=6$ : The four high-intensity regions begin to move towards the boundaries, with increasing heat accumulation at the corners. (f) $t=12$ : The quadrupole structure fades, and the heat current redistributes across the domain, reaching a more uniform state with peaks at the domain boundaries. This evolution highlights the transition from a concentrated central heat current to a more diffused distribution over time.}}
    \label{h2}
\end{figure*}

{\color{black}
The heat current results presented in \figref{h2} illustrate the time evolution of the numerical heat current, revealing a transition from a localized four-lobe pattern at early times to a more diffuse, boundary-influenced distribution at later times. The early-time four-lobe structure is not imposed by the square boundary itself. For a centered localized source, the early-time reduced envelope is approximately radial, $R_0(\vec{r},t)\approx R_0(r,t)$. The transverse heat current analyzed here is the $x$-component generated by a drive along $y$. In the Landau gauge, with $2e=2e$, the magnetic operator contains the mixed structure
$D_x
=
\partial_x+\frac{i 2e}{\hbar c}A_x
=
\partial_x-\frac{i 2eB}{\hbar c}y, A_x=-By$,
while the electric perturbation is proportional to $E_y y$. Acting on a centered localized profile, these two ingredients generate a leading anisotropic contribution of the schematic form $j_x^Q(\vec{r},t)\sim x\,y\,f(r,t)$, with $f(r,t)$ a radial function. This is the origin of the four off-center extrema observed at early times: the first anisotropic correction to the centered response is quadrupolar.

At early times $(t=0.1$ to $t=0.5)$, as shown in \figref{h2}(a)--(c), the heat current is concentrated in four symmetry-related regions around the origin. Around $t\sim 1$, the intensity further increases and the pattern expands outward. At later times, diffusion and boundary effects become increasingly important, so the redistribution toward the outer part of the domain should not be interpreted as universal geometry-independent behavior. The comparison between square and circular domains should therefore be read in a restricted sense: it shows that the early-time formation of the four-lobe structure is not a trivial artifact of one particular boundary, whereas the late-time redistribution is naturally more sensitive to finite boundaries, boundary treatment, and the detailed spatial profile of the initial excitation. We therefore interpret the quadrupolar pattern as a robust early-time precursor for the class of centered localized perturbations considered here, not as a universal late-time profile for arbitrary geometries or arbitrary initial conditions. For clarity, each panel in Figs.~6 and~7 uses its own color scale. Quantitative amplitude comparisons between different times should therefore be made from the reported peak values rather than by direct visual comparison of the color bars.
}

\vskip 2pt 

{\color{black}
We consider a representative NbN thin film~\cite{Chockalingam2008,Bouteiller2025,Beck2011} with $T_c\simeq 10~\mathrm{K}$, $\xi(0)\simeq 5~\mathrm{nm}$, and thickness $d\simeq 10~\mathrm{nm}$. For a perpendicular magnetic field $B=1~\mathrm{T}$, the Cooper-pair magnetic length is $\ell_B=\sqrt{\frac{\hbar}{2eB}}\approx 18.1~\mathrm{nm}$. Near the transition, the standard TDGL time scale is of order $\tau_{\mathrm{GL}}\sim \frac{\pi\hbar}{8k_B|T-T_c|},$ which is about $30~\mathrm{ps}$ for $|T-T_c|\approx 0.1~\mathrm{K}$. In the reduced Type-II model, $\tau_s$ should be interpreted as an effective local amplitude-recovery time of the same order, not as an exact microscopic vortex-transport time. The corresponding coherence length scale near the transition is $\xi(T)=\frac{\xi(0)}{\sqrt{|1-T/T_c|}}\approx 50~\mathrm{nm},$ so that the predicted textures live on spatial scales of roughly $20$--$50~\mathrm{nm}$ and temporal scales of a few tens of picoseconds. Experimentally, the most realistic protocol is a stroboscopic pump-probe measurement. A focused ultrafast optical or electrical pump can create a localized hot spot or partial local quench, and the subsequent recovery can then be reconstructed as a function of delay time by a phase-locked local probe. Picosecond superconducting recovery has already been resolved in NbN and related nitride films by time-resolved THz spectroscopy~\cite{Beck2011,Matsunaga2013,Matsunaga2014}. High-bandwidth scanning SQUID methods provide access to picosecond timing windows~\cite{Cui2017}, while SQUID-on-tip platforms and cryogenic magneto-thermoelectric imaging provide the relevant nanoscale local sensitivity~\cite{Zhou2023,Voelkl2024}. Achieving a full two-dimensional movie with both $\sim 20$--$50~\mathrm{nm}$ spatial resolution and $\sim 10$--$30~\mathrm{ps}$ temporal resolution remains challenging, so the predicted Type-II pattern should be viewed as a reproducible transient texture accessible by pump-probe averaging rather than as a routine single-shot image. The clearest practical signatures are likely to be: (i) the delay-dependent splitting of a central response into four off-center maxima, (ii) a transient overshoot of the off-axis signal below $T_c$, and (iii) the much stronger spatial inhomogeneity of the Type-II regime relative to the Type-I regime.
}

\section{Conclusion}\label{conc}
In summary, we find that the thermoelectric responses associated with Type-I and Type-II relaxation exhibit distinct dynamical characteristics. In the case of Type-I relaxation, which occurs when the temperature is above the critical temperature \(T_c\), the thermoelectric current reaches a steady state, it is spatially uniform and time-independent. This behavior arises from the predominance of linear relaxation mechanisms, which drive the system toward a homogeneous equilibrium configuration. In contrast, Type-II relaxation emerges at temperatures below \(T_c\), where nonlinear effects become dominant. As confirmed by numerical simulations, the resulting thermoelectric current is non-steady state: it exhibits pronounced spatial inhomogeneity and evolves dynamically over time. Initially, localized structures such as quadrupolar patterns may form; these structures subsequently diffuse and reorganize, giving rise to a temporally and spatially varying thermoelectric current. 

Moreover, the analytical and numerical findings of this study offer concrete theoretical model for designing and interpreting experiments \cite{behnia2015fundamentals, pourret2009nernst, behnia2016nernst} on superconducting thin films and mesoscopic devices under varied magnetic fields.
These qualitative differences in the thermal transport response provide clear, measurable signatures across $T_c$, With recent advances in local dissipation measurements—particularly the SQUID-on-tip technique \cite{zhou2023scanning, volkl2024demonstration}, which enables direct probing of spatiotemporal current distributions—such signatures should be experimentally accessible. This, in turn, offers a powerful diagnostic for distinguishing between the two relaxation regimes and for testing theoretical predictions concerning nonlinear saturation and characteristic crossover times. {\color{black}
Our Type-II results should therefore be understood as predictions for a short-time, local, amplitude-controlled nonlinear recovery regime below $T_c$, not as a complete theory of long-time vortex thermoelectric transport in the mixed state. Within this explicitly delimited scope, the contrast between the spatially uniform Type-I response and the transient quadrupolar Type-II response provides a concrete experimental discriminator across the superconducting transition.
} Looking ahead, the theoretical framework developed here may also prove valuable for exploring thermal transport phenomena in the quantum transport of topological superconducting systems \cite{li2024emergent,jiang2020geometric,jiang2022geometric}.

 \vskip 2pt

\section{Acknowledgments}
This work was financially supported by the National Natural Science Foundation of China (Grants No. 12374037), the Innovation
Program for Quantum Science and Technology Grant No.
2021ZD0301900, the National Key Research and Development Program of China (Grants No. 2024YFA1409001), and the Fundamental Research Funds for the Central Universities, Cultivation Project of
Shanghai Research Center for Quantum Sciences Grant
No. LZPY2024, and Shanghai Science and Technology Innovation Action Plan Grant No. 24LZ1400800.
\bibliography{Global_Note}

@article{Kazumi,
  title={Motion of the Vortex structure in Type-II Superconductors in high magnetic field},
  author={Caroli, C and Maki, K},
  journal={Physical Review },
  volume={164 },
  number={2 },
  pages={591-607 },
  year={1667},
  doi={https://doi.org/10.1103/PhysRev.164.591}
}

@article{rajeq,
  title={A New Numerical Method to Solve Some $PDE_s$  in the Unit Ball and Comparison with the Finite Element and the Exact Solution},
  author={Malek, R and  Ziti, C},
  journal={International Journal of Differential Equations.},
  volume={1 },
  pages={1-15 },
  year={2021},
doi={https://doi.org/10.1155/2021/6696165}
}

@article{Aa,
  title={A new entropic Riemann solver of conservation law mixed type including ziti's $\delta$-method with some experimental tests},
  author={Bsiss, L and Ziti, C},
  journal={Applied and Computational Mathematics},
  volume={6 },
  pages={222-232 },
  year={2017},
doi={10.11648/j.acm.20170605.12}
}

@article{Sco,
  title={Fluctuations near superconducting phase transitions},
  author={Skocpol, WJ and Tinkham, M},
  journal={Reports on Progress in Physics},
  volume={38},
  pages={21049-1097},
  year={1975},
doi={10.1088/0034-4885/38/9/001}
}

@article{Asla,
  title={The influence of fluctuation pairing of electrons on the conductivity of normal metal},
  author={Aslamasov, LG and Larkin, AI},
  journal={PHYSICS LETTERS},
  volume={26A, no 6},
  pages={238-239},
  year={1968},
doi={https://doi.org/10.1016/0375-9601(68)90623-3}
}

@book{Lar,
    author ={Larkin,A and Varlamov, A},
    title ={Theory of Fluctuations in Superconductors} ,
    publisher ={Clarendon Press},
    ISBN ={0-19-852815-9},
    year = {2005},
}

@article{Var,
  title={Fluctuation spectroscopy: From Rayleigh-Jeans waves to Abrikosov vortex clusters},
  author={Varlamov, AA and Galda, A and Glatz, A},
  journal={Reviews of Modern Physics},
  volume={90},
  pages={015009},
  year={2018},
doi={https://doi.org/10.1103/RevModPhys.90.015009}
}

@article{Ks,
  title={Singularity formation  in the Keller-Segel model: Theoretical analysis and numerical testing using the  $\delta_{-}$Ziti method},
  author={Chakir, S and Khalouq, S and Malek, R and Ziti, C},
  journal={International Journal of Applied and Computational Mathenatics},
  volume={11},
  pages={1-34},
  year={2025} ,
doi={https://doi.org/10.1007/s40819-025-01842-9}
}

@article{maki1968critical,
  title={Critical fluctuation of the order parameter in a superconductor. I},
  author={Maki, Kazumi},
  journal={Progress of Theoretical Physics},
  volume={40},
  number={2},
  pages={193--200},
  year={1968},
  publisher={Oxford University Press},
doi={https://doi.org/10.1143/PTP.40.193}
}

@article{huebener1995superconductors,
  title={Superconductors in a temperature gradient},
  author={Huebener, RP},
  journal={Superconductor Science and Technology},
  volume={8},
  number={4},
  pages={189},
  year={1995},
  publisher={IOP Publishing},
doi={10.1088/0953-2048/8/4/001}
}

@book{behnia2015fundamentals,
  title={Fundamentals of thermoelectricity},
  author={Behnia, Kamran},
  year={2015},
  publisher={OUP Oxford}
}

@article{pourret2009nernst,
  title={Nernst effect as a probe of superconducting fluctuations in disordered thin films},
  author={Pourret, A and Spathis, P and Aubin, H and Behnia, K},
  journal={New Journal of Physics},
  volume={11},
  number={5},
  pages={055071},
  year={2009},
  publisher={IOP Publishing},
doi={10.1088/1367-2630/11/5/055071}
}

@article{behnia2016nernst,
  title={Nernst effect in metals and superconductors: a review of concepts and experiments},
  author={Behnia, Kamran and Aubin, Herv{\'e}},
  journal={Reports on Progress in Physics},
  volume={79},
  number={4},
  pages={046502},
  year={2016},
  publisher={IOP Publishing},
doi={10.1088/0034-4885/79/4/046502}
}

@article{ullah1990critical,
  title={Critical fluctuations in high-temperature superconductors and the Ettingshausen effect},
  author={Ullah, Salman and Dorsey, Alan T},
  journal={Physical review letters},
  volume={65},
  number={16},
  pages={2066},
  year={1990},
  publisher={APS},
doi={https://doi.org/10.1103/PhysRevLett.65.2066}
}

@article{vishwanath2008nernst,
  title={Nernst effect and diamagnetism in phase fluctuating superconductors},
  author={Podolsky, Daniel and Raghu, Srinivas and Vishwanath, Ashvin},
  journal={Physical review letters},
  volume={99},
  number={11},
  pages={117004},
  year={2007},
  publisher={APS},
doi={https://doi.org/10.1103/PhysRevLett.99.117004}
}

@article{wang2006nernst,
  title={Nernst effect in high-T c superconductors},
  author={Wang, Yayu and Li, Lu and Ong, NP},
  journal={Physical Review B—Condensed Matter and Materials Physics},
  volume={73},
  number={2},
  pages={024510},
  year={2006},
  publisher={APS},
doi={https://doi.org/10.1103/PhysRevB.73.024510}
}

@article{cyr2018pseudogap,
  title={Pseudogap temperature T* of cuprate superconductors from the Nernst effect},
  author={Cyr-Choini{\`e}re, O and Daou, R and Lalibert{\'e}, F and Collignon, C and Badoux, Sven and LeBoeuf, D and Chang, J and Ramshaw, BJ and Bonn, DA and Hardy, WN and others},
  journal={Physical Review B},
  volume={97},
  number={6},
  pages={064502},
  year={2018},
  publisher={APS},
doi={https://doi.org/10.1103/PhysRevB.97.064502}
}

@article{hu2024vortex,
  title={Vortex entropy and superconducting fluctuations in ultrathin underdoped Bi2Sr2CaCu2O8+ x superconductor},
  author={Hu, Shuxu and Qiao, Jiabin and Gu, Genda and Xue, Qi-Kun and Zhang, Ding},
  journal={Nature Communications},
  volume={15},
  number={1},
  pages={4818},
  year={2024},
  publisher={Nature Publishing Group UK London},
doi={https://doi.org/10.1038/s41467-024-48899-6}
}

@article{schmid1966time,
  title={A time dependent Ginzburg-Landau equation and its application to the problem of resistivity in the mixed state},
  author={Schmid, Albert},
  journal={Physik der kondensierten Materie},
  volume={5},
  number={4},
  pages={302--317},
  year={1966},
  publisher={Springer},
doi={https://doi.org/10.1007/BF02422669}
}

@article{schmidt1968onset,
  title={The onset of superconductivity in the time dependent Ginzburg-Landau theory},
  author={Schmidt, Hartwig},
  journal={Zeitschrift f{\"u}r Physik A Hadrons and nuclei},
  volume={216},
  number={4},
  pages={336--345},
  year={1968},
  publisher={Springer},
doi={https://doi.org/10.1007/BF01391528}
}

@article{binder1973time,
  title={Time-dependent Ginzburg-Landau theory of nonequilibrium relaxation},
  author={Binder, K},
  journal={Physical Review B},
  volume={8},
  number={7},
  pages={3423},
  year={1973},
  publisher={APS},
doi={ https://doi.org/10.1103/PhysRevB.8.3423}
}

@article{tang1995time,
  title={Time dependent Ginzburg-Landau equations of superconductivity},
  author={Tang, Qi and Wang, S},
  journal={Physica D: Nonlinear Phenomena},
  volume={88},
  number={3-4},
  pages={139--166},
  year={1995},
  publisher={Elsevier},
doi={https://doi.org/10.1016/0167-2789(95)00195-A}
}

@article{rosenstein2010ginzburg,
  title={Ginzburg-Landau theory of type II superconductors in magnetic field},
  author={Rosenstein, Baruch and Li, Dingping},
  journal={Reviews of modern physics},
  volume={82},
  number={1},
  pages={109--168},
  year={2010},
  publisher={APS},
doi={https://doi.org/10.1103/RevModPhys.82.109}
}

@article{volkl2024demonstration,
  title={Demonstration and imaging of cryogenic magneto-thermoelectric cooling in a van der Waals semimetal},
  author={V{\"o}lkl, T and Aharon-Steinberg, A and Holder, T and Alpern, E and Banu, N and Pariari, AK and Myasoedov, Y and Huber, ME and H{\"u}cker, M and Zeldov, E},
  journal={Nature Physics},
  volume={20},
  number={6},
  pages={976--983},
  year={2024},
  publisher={Nature Publishing Group UK London},
  doi={https://doi.org/10.1038/s41567-024-02417-z}
}

@article{li2024emergent,
  title={Emergent energy dissipation in quantum limit},
  author={Li, Hailong and Jiang, Hua and Sun, Qing-Feng and Xie, XC},
  journal={Science Bulletin},
  volume={69},
  number={9},
  pages={1221--1227},
  year={2024},
  publisher={Elsevier},
  doi={https://doi.org/10.1016/j.scib.2024.03.024}
}

@article{jiang2022geometric,
  title={Geometric induction in chiral superfluids},
  author={Jiang, Qing-Dong and Balatsky, Alexander},
  journal={Physical Review Letters},
  volume={129},
  number={1},
  pages={016801},
  year={2022},
  publisher={APS}
}

@article{jiang2020geometric,
  title={Geometric induction in chiral superconductors},
  author={Jiang, Qing-Dong and Hansson, Thors Hans and Wilczek, Frank},
  journal={Physical Review Letters},
  volume={124},
  number={19},
  pages={197001},
  year={2020},
  publisher={APS}
}

@article{zhou2023scanning,
  title={Scanning SQUID-on-tip microscope in a top-loading cryogen-free dilution refrigerator},
  author={Zhou, Haibiao and Auerbach, Nadav and Roy, Indranil and Bocarsly, Matan and Huber, Martin E and Barick, Barun and Pariari, Arnab and H{\"u}cker, Markus and Lim, Zhi Shiuh and Ariando, A and others},
  journal={Review of scientific instruments},
  volume={94},
  number={5},
  year={2023},
  publisher={AIP Publishing}
}

@article{Ussishkin2002PRL,
  author    = {I. Ussishkin and S. L. Sondhi and D. A. Huse},
  title     = {Gaussian Superconducting Fluctuations, Thermal Transport, and the Nernst Effect},
  journal   = {Physical Review Letters},
  volume    = {89},
  number    = {28},
  pages     = {287001},
  year      = {2002},
  doi       = {10.1103/PhysRevLett.89.287001},
  url       = {https://doi.org/10.1103/PhysRevLett.89.287001}
}

@article{Bouteiller2025,
  author  = {Bouteiller, T. and Marguerite, A. and Daou, R. and Yakovlev, D. and Pons, S. and Feuillet-Palma, C. and Roditchev, D. and Fauqu{\'e}, B. and Behnia, K.},
  title   = {Nernst effect and its thickness dependence in superconducting NbN films},
  journal = {Physical Review Materials},
  volume  = {9},
  pages   = {094807},
  year    = {2025},
  doi     = {10.1103/PhysRevMaterials.9.094807}
}

@article{Shokri2025,
  author  = {Shokri, S. and Ceccardi, M. and Confalone, T. and Saggau, C. N. and Lee, Y. and Martini, M. and Gu, G. and Vinokur, V. M. and Pallecchi, I. and Nielsch, K. and Caglieris, F. and Poccia, N.},
  title   = {Evolution of Dissipative Regimes in Atomically Thin Bi$_2$Sr$_2$CaCu$_2$O$_{8+x}$ Superconductor},
  journal = {Advanced Electronic Materials},
  volume  = {11},
  pages   = {2400496},
  year    = {2025},
  doi     = {10.1002/aelm.202400496}
}

@article{Schmid1966,
  author  = {A. Schmid},
  title   = {A time dependent Ginzburg-Landau equation and its application to the problem of resistivity in the mixed state},
  journal = {Phys. Kondens. Mater.},
  volume  = {5},
  pages   = {302--317},
  year    = {1966},
  doi     = {10.1007/BF02422669}
}

@book{LarkinVarlamov2005,
  author    = {A. I. Larkin and A. A. Varlamov},
  title     = {Theory of Fluctuations in Superconductors},
  publisher = {Oxford University Press},
  address   = {Oxford},
  year      = {2005}
}

@article{Rosenstein2010,
  author  = {B. Rosenstein and D. Li},
  title   = {Ginzburg-Landau theory of type II superconductors in magnetic field},
  journal = {Rev. Mod. Phys.},
  volume  = {82},
  pages   = {109--168},
  year    = {2010},
  doi     = {10.1103/RevModPhys.82.109}
}

@article{CaroliMaki1967,
  author  = {C. Caroli and K. Maki},
  title   = {Motion of the Vortex Structure in Type-II Superconductors in High Magnetic Field},
  journal = {Phys. Rev.},
  volume  = {164},
  number  = {2},
  pages   = {591--607},
  year    = {1967},
  doi     = {10.1103/PhysRev.164.591}
}

@article{Podolsky2007,
  author  = {D. Podolsky and S. Raghu and A. Vishwanath},
  title   = {Nernst Effect and Diamagnetism in Phase Fluctuating Superconductors},
  journal = {Phys. Rev. Lett.},
  volume  = {99},
  pages   = {117004},
  year    = {2007},
  doi     = {10.1103/PhysRevLett.99.117004}
}

@article{BehniaAubin2016,
  author  = {K. Behnia and H. Aubin},
  title   = {Nernst effect in metals and superconductors: a review of concepts and experiments},
  journal = {Rep. Prog. Phys.},
  volume  = {79},
  pages   = {046502},
  year    = {2016},
  doi     = {10.1088/0034-4885/79/4/046502}
}

@article{Hu2024,
  author  = {S. Hu and J. Qiao and G. Gu and Q.-K. Xue and D. Zhang},
  title   = {Vortex entropy and superconducting fluctuations in ultrathin underdoped Bi$_2$Sr$_2$CaCu$_2$O$_{8+x}$ superconductor},
  journal = {Nat. Commun.},
  volume  = {15},
  pages   = {4818},
  year    = {2024},
  doi     = {10.1038/s41467-024-48899-6}
}

@article{Wu2024,
  author  = {Y. Wu and A. Roy and S. Dutta and J. Jesudasan and P. Raychaudhuri and A. Frydman},
  title   = {Nernst Sign Reversal in the Hexatic Vortex Phase of Weakly Disordered a-MoGe Thin Films},
  journal = {Phys. Rev. Lett.},
  volume  = {132},
  number  = {2},
  pages   = {026003},
  year    = {2024},
  doi     = {10.1103/PhysRevLett.132.026003}
}

@article{Chockalingam2008,
  author  = {S. P. Chockalingam and M. Chand and J. Jesudasan and V. Tripathi and P. Raychaudhuri},
  title   = {Superconducting properties and Hall effect of epitaxial NbN thin films},
  journal = {Phys. Rev. B},
  volume  = {77},
  pages   = {214503},
  year    = {2008},
  doi     = {10.1103/PhysRevB.77.214503}
}

@article{Beck2011,
  author  = {M. Beck and M. Klammer and S. Lang and P. Leiderer and V. V. Kabanov and G. N. Gol'tsman and J. Demsar},
  title   = {Energy-Gap Dynamics of Superconducting NbN Thin Films Studied by Time-Resolved Terahertz Spectroscopy},
  journal = {Phys. Rev. Lett.},
  volume  = {107},
  pages   = {177007},
  year    = {2011},
  doi     = {10.1103/PhysRevLett.107.177007}
}

@article{Matsunaga2013,
  author  = {R. Matsunaga and Y. I. Hamada and K. Makise and Y. Uzawa and H. Terai and Z. Wang and R. Shimano},
  title   = {Higgs Amplitude Mode in the BCS Superconductors Nb$_{1-x}$Ti$_x$N Induced by Terahertz Pulse Excitation},
  journal = {Phys. Rev. Lett.},
  volume  = {111},
  pages   = {057002},
  year    = {2013},
  doi     = {10.1103/PhysRevLett.111.057002}
}

@article{Matsunaga2014,
  author  = {R. Matsunaga and N. Tsuji and H. Fujita and A. Sugioka and K. Makise and Y. Uzawa and H. Terai and Z. Wang and H. Aoki and R. Shimano},
  title   = {Light-induced collective pseudospin precession resonating with Higgs mode in a superconductor},
  journal = {Science},
  volume  = {345},
  number  = {6201},
  pages   = {1145--1149},
  year    = {2014},
  doi     = {10.1126/science.1254697}
}

@article{Cui2017,
  author  = {Z. Cui and J. R. Kirtley and Y. Wang and P. A. Kratz and A. J. Rosenberg and C. A. Watson and G. W. Gibson and M. B. Ketchen and K. A. Moler},
  title   = {Scanning SQUID sampler with 40-ps time resolution},
  journal = {Rev. Sci. Instrum.},
  volume  = {88},
  number  = {8},
  pages   = {083703},
  year    = {2017},
  doi     = {10.1063/1.4986525}
}

@article{Zhou2023,
  author  = {H. Zhou and N. Auerbach and I. Roy and M. Bocarsly and M. E. Huber and B. Barick and A. Pariari and M. H\"ucker and Z. S. Lim and A. Ariando and A. I. Berdyugin and N. Xin and M. Rappaport and Y. Myasoedov and E. Zeldov},
  title   = {Scanning SQUID-on-tip microscope in a top-loading cryogen-free dilution refrigerator},
  journal = {Rev. Sci. Instrum.},
  volume  = {94},
  number  = {5},
  pages   = {053706},
  year    = {2023},
  doi     = {10.1063/5.0142073}
}

@article{Voelkl2024,
  author  = {T. V\"olkl and A. Aharon-Steinberg and T. Holder and E. Alpern and N. Banu and A. K. Pariari and Y. Myasoedov and M. E. Huber and M. H\"ucker and E. Zeldov},
  title   = {Demonstration and imaging of cryogenic magneto-thermoelectric cooling in a van der Waals semimetal},
  journal = {Nat. Phys.},
  volume  = {20},
  number  = {6},
  pages   = {976--983},
  year    = {2024},
  doi     = {10.1038/s41567-024-02417-z}
}

  \appendix
  \onecolumngrid


\section{Overview of the $\delta_{-}$Ziti method} \label{numapp}
For completeness and ease of reading, we present a self-contained overview of the numerical method used in this paper. The $\delta_{-}$Ziti method has proven effective in handling strongly nonlinear PDEs across various contexts \cite{rajeq, Aa, Ks}. A key advantage of this approach is its ability to capture transitions between regular and singular behaviors without requiring finer mesh refinement, making it particularly suitable for relaxation equations with nonlinear terms. By employing the $\delta_{-}$Ziti method, we extend its applicability to new domains while leveraging its accuracy in modeling nonlinear dynamics. Mathematically, the method is based on the classical Galerkin variational formulation, with its core step being the construction of an orthonormal basis in one dimension. This is achieved using the well-known function ${\mathcal{F}}$ defined for any $x \in \mathbb{R}$ as:

\begin{equation}
 \mathcal{F}(x)=
\begin{cases}
{\exp\left(\frac{1}{\lvert x\lvert^2 -1}\right)} & \text{if $\lvert x \rvert < 1$} \\
{0}  & \text{otherwise.}
\end{cases} 
\label{Phi}
\end{equation}

Below, we outline the key steps involved in constructing the basis functions in the one-dimensional case.
\subsection{Construction of the basis functions of $\delta_{-}$Ziti method}
 Let $I=[a, b]$ be a given interval. For an integer $m>0$, we define the spatial step size as $h=\frac{b-a}{m}$, with nodes given by
$$x_i=a+(i-1)h, \ \forall i \in \{1, \cdots,m+1\}.$$ 
We them construct the family  $(\varphi_i)$ as follows:
\begin{equation*}
\varphi_i(x):=\frac{C}{h}{\mathcal{F}}\left(\frac{x-x_i}{h}\right)~~~~  \ \text{for} \ \  i \in\{1,\cdots,m+1\} \,,
\end{equation*}
where  $C$ is the normalization constant:
$$C:=  \displaystyle \frac{1}{ \displaystyle \int_{\mathbb{R}} {\mathcal{F}}(x)dx}.$$ \\
The support of each function is given by: 
\begin{equation*}
 \operatorname{supp}\left(\varphi_i\right)= \begin{cases}{\left[x_1, x_2\right],} & i=1 \\ {\left[x_{i-1}, x_{i+1}\right],} & 2 \leq i \leq m \\ {\left[x_m, x_{m+1}\right],} & i=m+1.\end{cases}   
\end{equation*}
Moreover, the set  $\left(\varphi_i \right)_{1\leq i \leq m+1}$ is linearly independent in  $L^2(\rm I)$. Using the Gram-Schmidt process, we construct the orthogonal family, denoted $ \left( \tilde{\psi }_i \right)$, satisfying the following relation
\begin{equation}
\left\lbrace
\begin{aligned} \label{psi}
& \tilde{\psi}_{1}(x) = \varphi_{1}(x),\\
& \tilde{\psi}_{i}(x) = \varphi_{i}(x)+\lambda_{i-1} \tilde{\psi}_{i-1}(x),  \quad 2\leq i\leq m+1,\\
\end{aligned}
\right. 
\end{equation}
where 
\begin{equation*}
\lambda_{i-1}=-\frac{\left\langle\varphi_i, \tilde{\psi}_{i-1}\right\rangle}{\left\langle\tilde{\psi}_{i-1}, \tilde{\psi}_{i-1}\right\rangle} .
\end{equation*}
Then, we  normalize  $\tilde{\psi}_{i}(x)$ to obtain the final orthonormal basis:
$$\psi_i(x)= \displaystyle \frac{\tilde{\psi}_{i}(x)}{|| \tilde{\psi_i}||_{L^2(\rm I)}}.$$\\

\subsection{The $\delta_{-}$Ziti approximation relations}
The aim of  $\delta_{-}$ziti approximation involves projecting a given function 
$f$ onto the finite-dimensional space spanned by the $\left(\psi_i\right)$, and approximating its integral accordingly. This gives
\begin{equation}
 f(x)  \simeq \sum_{i=1}^{m+1} c_i \psi_{i}(x),\label{12}   
\end{equation}
 where
 \begin{equation}
     c_i =  \int_{a}^{b} f(x)  \psi_{i}(x)dx. \label{13}
 \end{equation}
A key property of this method is its ability to express integrals in terms of the roots $r_i$ of the basis functions $\psi_i$ :

\begin{align}
&\  c_i \approx \frac{f\left(r_i\right)}{\psi_i\left(r_i\right)} \label{app2}\\
&\int_a^b f(x) \psi_i(x) d x \approx \frac{f\left(r_i\right)}{\psi_i\left(r_i\right)} \label{app3}
\end{align}

These relations are crucial for constructing the numerical scheme associated with our system. The above one-dimensional framework generalizes to higher dimensions via tensorization. 
\subsection{Construction of the $\delta_{-}$ Ziti scheme}
Now, we are ready to outline the  important steps in the construction of  $\delta_{-}$ziti scheme. 
From the formula  Eq.~\eqref{12}, we can approximate the propagator as follows: 

\begin{equation}
 R_0(\vec{r},t;\vec{r_0};t_0)   \simeq \sum^{m+1}_{\ell,p=1} a_{\ell p}(t) \psi_{\ell p}(\vec{r})  \label{ap1}\\
\end{equation}
where \begin{equation*}
\forall \vec{r}=(x,y) \in \rm\Omega=]0,1[^2, \quad 	\psi_{\ell p}(\vec{r})= \psi_{\ell}(\it x) \psi_{p}( \it y).
\end{equation*}
The numerical scheme is constructed following these key steps:

\begin{enumerate}
\item[(i)] \textbf{Galerkin Projection:} Multiply the given equation by a basis function and integrate over the domain \( \Omega \), following the Galerkin method.

\item[(ii)] \textbf{Application of the Approximation Formula:} Apply the approximation formula Eq.~ \eqref{app3} as it stands.

\item[(iii)] \textbf{Substituting the Approximation Formula:} Injecting the approximation formula Eq.~\eqref{ap1} into the resulting equation from the previous step, we obtain:
\begin{align}
\begin{split}
\int_{\Omega} \frac{\partial R_0}{\partial t}(t,\vec{r}) \psi_{ij}(\vec{r}) d\vec{r}  
&\simeq \int_{\Omega} \sum_{\ell,p=1}^{m+1} a_{\ell p }^{\prime}(t) \psi_{\ell p }(\vec{r}) \psi_{ij}(\vec{r}) d\vec{r} \\  
&=\sum_{\ell,p=1}^{m+1} a_{\ell p}^{\prime}(t) \int_{\Omega} \psi_{\ell p}(\vec{r}) \psi_{ij}(\vec{r}) d\vec{r} \\  
&= a_{ij}^{\prime}(t).
\label{a'}
\end{split}
\end{align}

\item[(iv)] \textbf{Temporal Discretization:} The time interval \( [0,T] \) is divided into \( N \) uniform sub-intervals \( [t^n, t^{n+1}] \) with a time step \( \Delta t = t^{n+1} - t^n \) for \( n = 0, \dots, N-1 \). Denoting \( a^n_{ij} \) as the approximation of \( a_{ij}(t) \) at \( t = t^n \), we use a semi-discrete time approximation:
\begin{equation}
a_{ij}^{\prime}(t) \simeq \frac{a_{ij}^{n+1} - a_{ij}^{n}}{\Delta t}.
\label{p1}
\end{equation}
\end{enumerate}
For the remaining terms, we extend the use of formula Eq.~\eqref{app3} to the bi-dimensional case, allowing us to approximate each term and construct the final numerical scheme. \vskip 2pt

{\color{black}Unlike standard numerical methods such as the Crank–Nicolson or explicit Euler schemes, the $\delta_{-}$Ziti approach constructs an orthonormal basis that directly incorporates both spatial diffusion and temporal relaxation.  This Galerkin-type formulation allows stable integration of both linear and strongly nonlinear terms without requiring iterative updates or  matrix inversion at each time step, as in implicit schemes. Moreover, the $\delta_{-}$Ziti basis ensures local support and rapid convergence,  enabling accurate treatment of sharp gradients and dissipative relaxation dynamics with relatively coarse grids. 
These features make the method particularly efficient for time-dependent Ginzburg–Landau equations, where it maintains stability and accuracy in regimes that challenge conventional solvers.}

\section{Robustness of the Early-Time Four-Lobe Heat-Current Pattern Across Domain Shapes}

\begin{figure*}[!htbp] 
    \centering  
    \minipage{0.33\textwidth}
    \includegraphics[height=1.4in, width=1\linewidth]{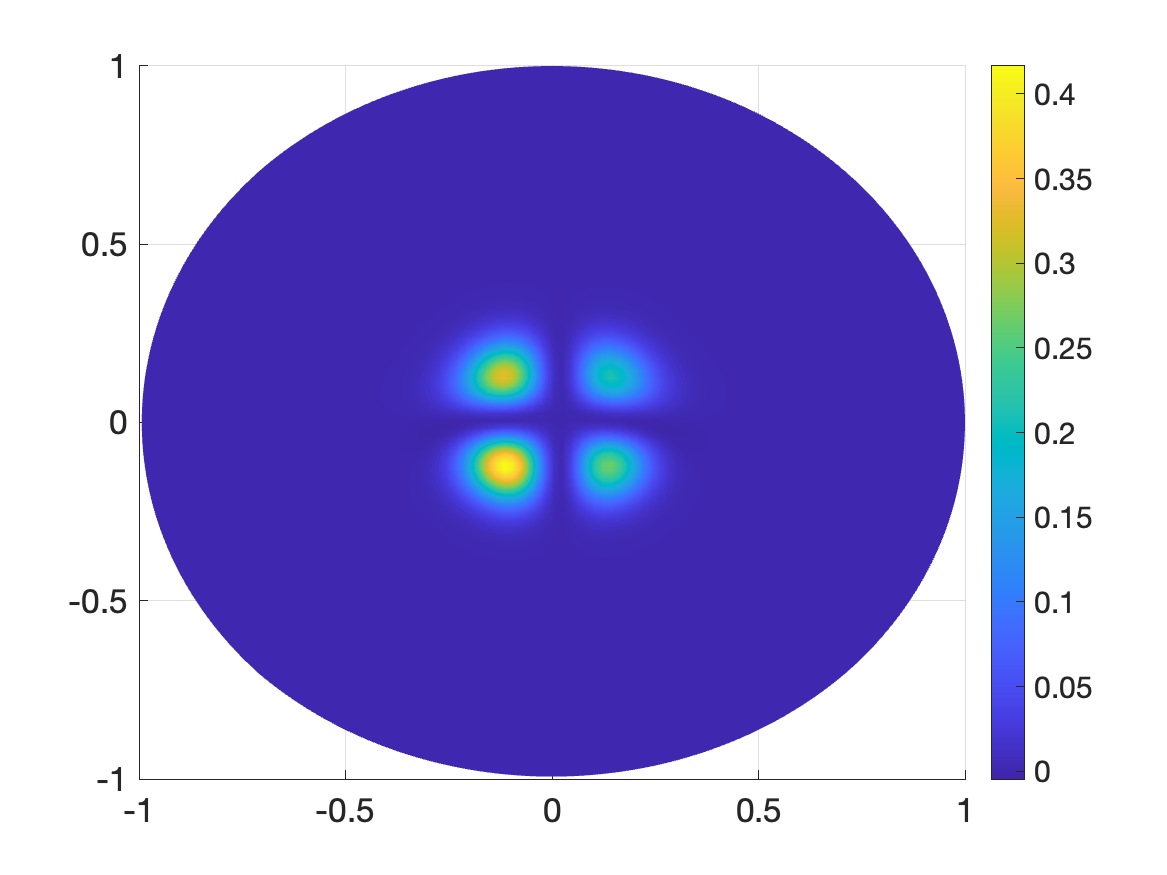}
        \subcaption{$t=0.1$}
    \endminipage
    \minipage{0.33\textwidth}
        \includegraphics[height=1.4in, width=2.3in]{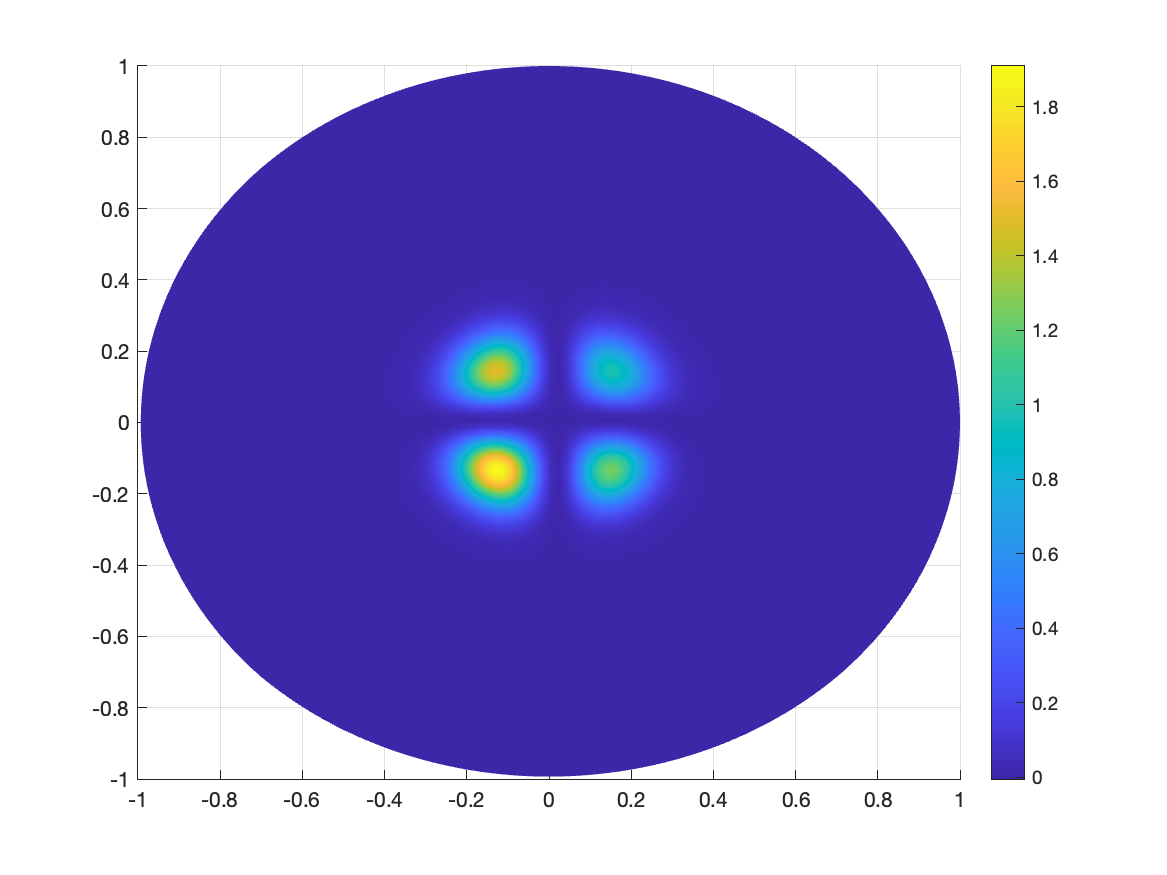}
        \subcaption{$t=0.2$}
    \endminipage
    \minipage{0.33\textwidth}
        \includegraphics[height=1.4in, width=2.3in]{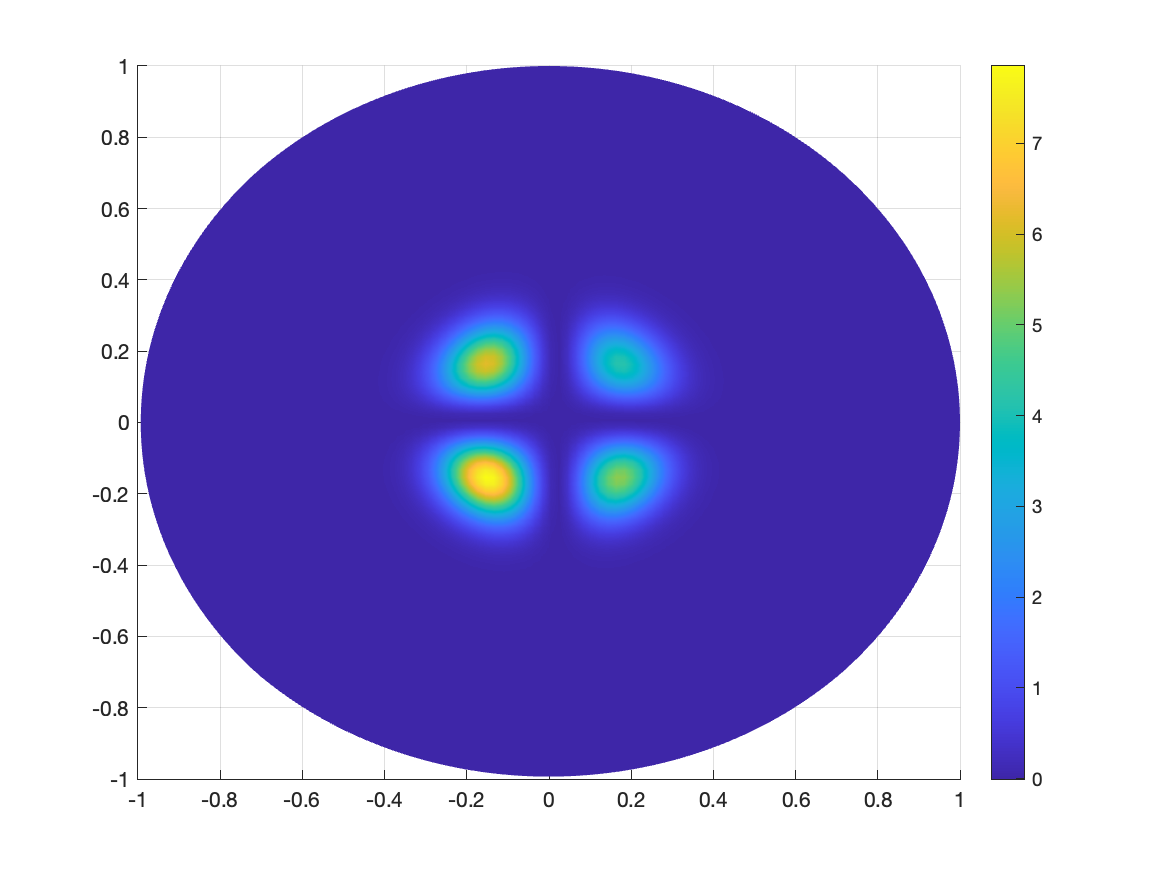}
        \subcaption{$t=0.5$}
    \endminipage
    \\ 
    \minipage{0.33\textwidth}
        \includegraphics[height=1.4in, width=2.3in]{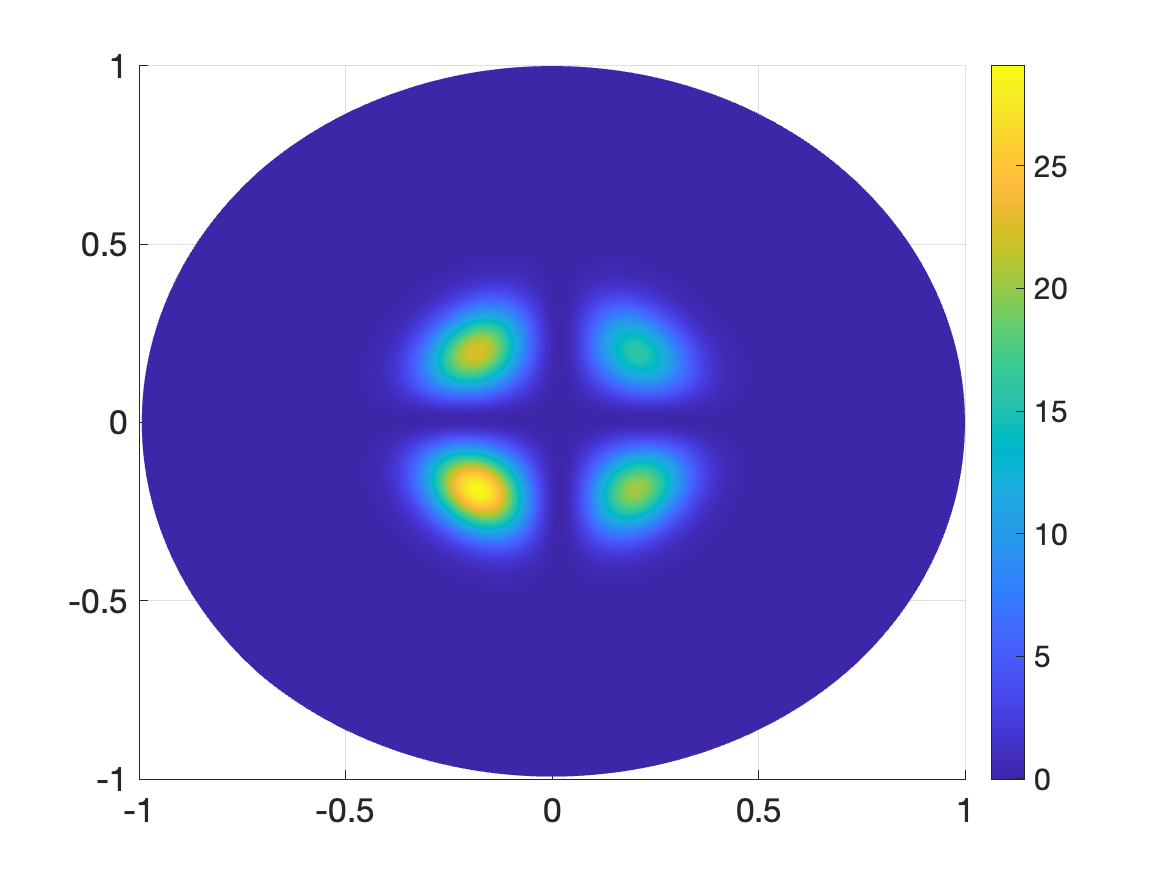}
        \subcaption{$t=1.0$}
    \endminipage
    \minipage{0.33\textwidth}
        \includegraphics[height=1.4in, width=2.3in]{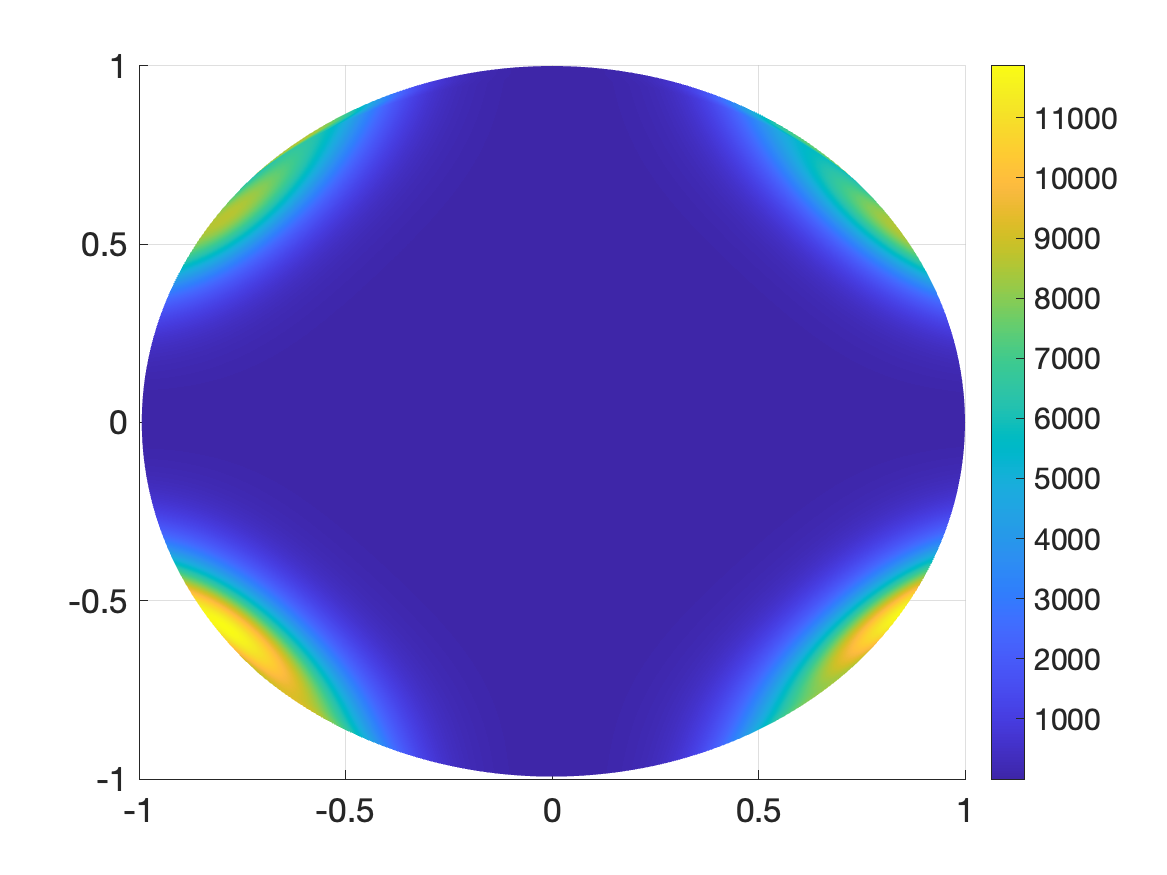}
        \subcaption{$t=6.0$}
    \endminipage
    \minipage{0.33\textwidth}
        \includegraphics[height=1.4in, width=2.3in]{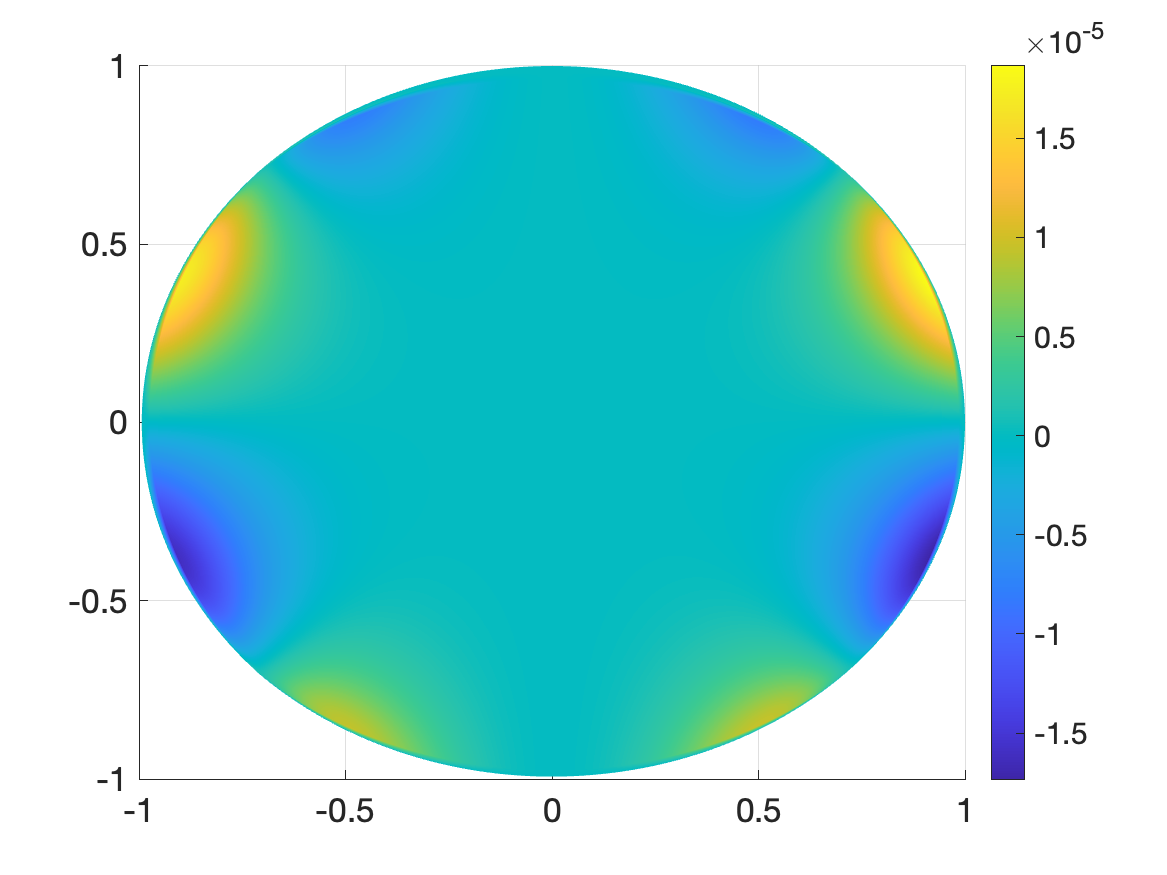}
        \subcaption{$t=12.0$}
    \endminipage
 \caption{\justifying {The numerical heat current at different time. The figure shows the numerical heat current in a circular domain at various time steps, labeled (a) through (f). (a) $t=0.1$: A quadrupole structure forms near the center with four localized high-intensity regions. (b) $t=0.2$ : The peak intensity increases while the quadrupolar pattern remains confined near the origin. (c) $t=0.5$ : The heat current continues to grow in magnitude, preserving the fourfold structure. (d) $t=1$ : The high-intensity regions expand outward, indicating the onset of nonlinear energy redistribution. (e) $t=6$ : The heat current spreads toward the outer region of the disc, with maxima shifting closer to the periphery. (f) $t=12$ : The quadrupolar symmetry weakens, and the current evolves into a smoother, nearly uniform profile across the domain. This sequence shows that the early-time four-lobe precursor is robust across square and circular domains for the same centered localized excitation. By contrast, the later-time redistribution remains sensitive to finite boundaries and should not be interpreted as fully geometry independent.}}
    \label{fig:disc}
\end{figure*}

{\color{black}
To test whether the early-time four-lobe structure observed in \figref{h2} is imposed by the square computational box, we numerically solve the same evolution equation in a circular region. The physical parameters, centered localized excitation, and numerical discretization scheme are chosen to match those used for the square domain as closely as possible. \figref{fig:disc} displays the time evolution of the heat-current intensity for representative times $t=0.1,0.2,0.5,1.0,6.0,$ and $12.0$. At early stages $(t\leq 0.5)$, four localized maxima emerge near the center, forming a quadrupolar pattern similar to that observed in the square domain. At later times, diffusion and finite-boundary effects become increasingly important, and the heat current redistributes over the entire domain.
}

{\color{black}
The circular-domain calculation should therefore be interpreted in a restricted sense. Numerically, the appearance of the same four-lobe structure at early times in two different domain shapes supports the claim that the early-time quadrupolar precursor is not a trivial artifact of the square geometry. Physically, this behavior originates from the interplay between diffusion, magnetic coupling, and nonlinear saturation for a centered localized excitation. By contrast, the late-time redistribution is naturally more sensitive to finite boundaries and to the detailed spatial profile of the initial condition. The circular geometry thus provides a complementary demonstration of the robustness of the early-time four-lobe pattern, rather than a proof of fully geometry-independent dynamics at all times.
}

\end{document}